\documentclass[preprint,12pt]{elsarticle}

\usepackage{graphicx}

\usepackage{amssymb}
\usepackage{amsmath}

\biboptions{sort&compress}

\begin{document}

\begin{frontmatter}



\title{Chaos-assisted formation of immiscible matter-wave solitons and self-stabilization in 
the binary discrete nonlinear Schr\"odinger equation}


\author{D.V.~Makarov}
\author{M.Yu.~Uleysky}

\address{Laboratory of Nonlinear Dynamical Systems,
V.I.~Il'ichev Pacific Oceanological Institute of the Far-Eastern Branch of the Russian Academy of Sciences,\\
43 Baltiiskaya st., 690041 Vladivostok, Russia, URL: http://dynalab.poi.dvo.ru}

\begin{abstract}

Binary discrete nonlinear Schr\"odinger equation is used to describe dynamics
of two-species Bose-Einstein condensate loaded into an optical lattice. 
Linear inter-species coupling leads to Rabi transitions between the species.
In the regime of strong nonlinearity, a wavepacket corresponding to condensate 
separates into localized and ballistic fractions.
Localized fraction is predominantly formed by immiscible solitons consisted of only one species. 
Initial states without spatial separation of occupied sites expose formation
of immiscible solitons after a strongly chaotic transient.
We calculate the finite-time Lyapunov exponent as a rate of wavepacket divergence in the Hilbert space.
Using the Lyapunov analysis supplemented by Monte-Carlo sampling, 
it is shown that appearance of immiscible solitons after the chaotic
transient corresponds to self-stabilization of the wavepacket.
It is found that onset of chaos is accompanied by fast variations of kinetic and 
interaction energies.
Crossover to self-stabilization is accompanied by reduction of condensate density due to
emittance of ballistically propagating waves.
It turns out that spatial separation of species should be a necessary condition
for wavepacket stability in the presence of linear inter-species coupling.

\end{abstract}


\end{frontmatter}



\section{Introduction}
\label{intro}

Discrete nonlinear Schr\"odinger equation (DNLSE) being coupled system
of nonlinear ordinary differential equations occurs in variety of physical problems \cite{Kevrekidis-survey}.
For example, DNLSE describes light propagation in periodic waveguide arrays.
If the array consists of two distinct chains coupled to each other, 
then a light wavefield is governed by the binary DNLSE \cite{Locatelli04}.
Recently such configuration was exploited for optical simulation of neutrino oscillations,
and formation of neutrino solitons was predicted \cite{Neutrino-PRL14}.
In quantum optics, DNLSE is utilized for studying Bose-Einstein condensate (BEC)
loaded into an optical lattice.
Binary DNLSE occurs if the condensate is a mixture of two different atomic species.
The species may correspond to different atoms, for example, $^{133}$Cs and
$^{87}$Rb \cite{Lercher}. Another possibility is the mixture of atoms of the same sort but belonging
to different hyperfine states \cite{Vidanovic}. In the latter case the species can be transformed
to each other by means of external electromagnetic field giving rise to resonant 
transitions between the hyperfine states.

One of the main properties of the DNLSE is the onset
of discrete solitons and/or breathers when the nonlinearity is strong enough \cite{Flach_Willis,Gorbach06,Franzosi}.
Under certain conditions, their formation can be studied within the variational approach that relies
upon a priori assumptions about the wavepacket form in course of its evolution \cite{TrSmerzi}.
However, as it was found in a recent work \cite{JRLR14},
solitons in the binary DNLSE can arise spontaneously after a transient regime of highly irregular dynamics,
when the variational approach doesn't apply.
A distinctive feature of such solitons is that they consist of only one species
and well separated from each other. In the case of BEC mixtures loaded into optical lattices,
onset of one-species (i.e., immiscible) solitons corresponds to the miscible-immiscible
quantum phase transition \cite{Merhasin_Malomed_Driben,Nicklas-PRL11,FeiZhan-miscible} being a form 
of spontaneous symmetry breaking \cite{Gubeskys,Adhikari_Malomed-PRA09}.

In the present work,
we examine the link between the spontaneous formation of matter-wave solitons and Lyapunov stability.
Our attention is concentrated on the regime of strong nonlinearity
that corresponds to more extensive interaction between atoms and anticipates more pronounced manifestation
of chaos.

The paper is organized as follows. In the next section we describe the model under consideration.
Section \ref{Undistort} represents some generic features of condensate chaotic dynamics and soliton formation.
The Monte-Carlo method is utilized for studying various aspects of wavepacket dynamics
in Section \ref{Monte-Carlo}. Section \ref{Summ} is devoted to discussion of the main results and outlines some
prospects for further research.

\section{Basic equations}
\label{Basic}

Binary DNLSE is given by the set of coupled ODE
\begin{equation}
\begin{aligned}
 i\hbar\frac{da_n}{dt}&=-\frac{J}{2}(a_{n-1} + a_{n+1}) + g_{11}|a_n|^2a_n + g_{12}|b_n|^2a_n -\frac{\hbar\Omega}{2}b_n,\\
 i\hbar\frac{db_n}{dt}&=-\frac{J}{2}(b_{n-1} + b_{n+1}) + g_{22}|b_n|^2b_n + g_{21}|a_n|^2b_n -\frac{\hbar\Omega}{2}a_n.
 \end{aligned}
\label{TB}
\end{equation}
Our interest is mainly motivated by the problem of BEC dynamics in optical lattices. 
In this way, Eqs. (\ref{TB}) describe motion of two-species BEC mixture, where the species
correspond to different hyperfine states. 
Quantities $a_n$ and $b_n$ are complex-valued amplitudes of BEC wave function at the lattice site $n$.
$J$ is a hopping rate quantifying tunneling between neighbouring sites. $\Omega$ 
is a Rabi frequency being the rate of inter-species transitions between 
the hyperfine states under the action of the external rf radiation.
Nonlinearity coefficients $g_{ij}$ 
are determined by $s$-wave scattering lengths and
quantify strength of interaction between atoms.
As in Ref.~\cite{JRLR14}, we assume that the interaction between atoms belonging to different hyperfine states
is weak and can be neglected, $g_{12}=g_{21}=0$.
Indeed, both intra-species and inter-species scattering lengths 
can be readily tuned in experiments 
by means of the Feshbach resonance \cite{Turlapov-JETPL-eng,Lozada-Vera}, that is, one has some freedom in choosing their values.
Also, it should be noted that intra-species dynamics rather depends 
on the difference of nonlinearity parameters
\begin{equation}
 \Delta g = g_{11} + g_{22} - g_{12}- g_{21},
\end{equation}
than on their absolute calues \cite{Leggett,Nicklas-PRL11},
therefore one can regard the case of $g_{12}=g_{21}=0$ ($g_{11,22}\ne 0$) as the limiting regime when 
intra-species interaction dominates.
As both species correspond to same sort of atoms, one can set $g_{11}=g_{22}\equiv g$.
Thus, we arrive to a simplified form of (\ref{TB})
\begin{equation}
\begin{aligned}
 i\hbar\frac{da_n}{dt}&=-\frac{J}{2}(a_{n-1} + a_{n+1}) + g|a_n|^2a_n -\frac{\hbar\Omega}{2}b_n,\\
 i\hbar\frac{db_n}{dt}&=-\frac{J}{2}(b_{n-1} + b_{n+1}) + g|b_n|^2b_n -\frac{\hbar\Omega}{2}a_n,
 \end{aligned}
\label{TB2}
\end{equation}
Equations (\ref{TB2}) can be also considered as the semiclassical limit of the corresponding 
two-species Bose-Hubbard model \cite{FeiZhan-miscible}.
Hereafter we shall use the normalization condition
\begin{equation}
 \sum\limits_{n=-N}^{N} \rho_n = 1,\quad
 \rho_n = |a_n|^2 + |b_n|^2,
 \label{norm}
\end{equation}
where $\rho_n$ is population of the site $n$, $2N+1$ is total number of sites.
Amplitudes $a_n$ and $b_n$ can be represented as
\begin{equation}
a_n = \sqrt{A_n}e^{i\alpha_n},\quad
b_n = \sqrt{B_n}e^{i\beta_n},\quad
A_n,B_n,\alpha_n,\beta_n\in \Re.
 \label{amplrepr}
\end{equation}
%
Substituting (\ref{amplrepr}) into (\ref{TB}),
we can rewrite (\ref{TB}) in the Hamiltonian form
\begin{equation}
\begin{aligned}
 \frac{dA_n}{dt}&=-\frac{\partial H}{\partial\alpha_n},\quad
 \frac{d\alpha_n}{dt}=\frac{\partial H}{\partial A_n},\\
 \frac{dB_n}{dt}&=-\frac{\partial H}{\partial\beta_n},\quad
 \frac{d\beta_n}{dt}=\frac{\partial H}{\partial B_n}.
\end{aligned}
\label{Hsys}
\end{equation}
The corresponding Hamiltonian is given by the sum
\begin{equation}
H = H_{\text{kin}}+H_{\text{Rabi}}+H_{\text{int}},
 \label{Ham0}
\end{equation}
where 
\begin{equation}
H_{\text{kin}} = -J\sum\limits_{n=-N}^{N-1}[
\sqrt{A_nA_{n+1}}\cos(\alpha_n-\alpha_{n+1}) + 
\sqrt{B_nB_{n+1}}\cos(\beta_n-\beta_{n+1})]
\label{Ekin}
\end{equation}
is the kinetic energy,
\begin{equation}
 H_{\text{Rabi}}=-\Omega\sum\limits_{n=-N}^N\sqrt{A_nB_n}\cos(\alpha_n-\beta_n)
 \label{ERabi}
\end{equation}
is the energy of inter-species coupling, or Rabi energy,
\begin{equation}
H_{\text{int}}=\frac{g}{2}\sum\limits_{n=-N}^N(A_n^2 + B_n^2)
\label{Eint}
\end{equation}
is the energy of interaction between atoms belonging to the same species.
Formally, the Hamiltonian described by (\ref{Ham0})--(\ref{Eint})
looks as Hamiltonian of a binary chain of classical particles with nearest-neighbour
coupling that depends on $A_n$ and $B_n$  playing the role of ``particle velocities''.
Continuing this analogy,
the term $H_{\text{int}}$ enters into the Hamiltonian as the ``particle kinetic energy'',
and $g$ is the inverse ``mass''. 
One can easily see that in the limit $J,\Omega \to 0$ the equations (\ref{Hsys}) become
integrable. This case corresponds to formation of one-site standing solitons, also known as compactons 
\cite{FAbdullaev-cton10,FAbdullaev-cton14,Bazeia}.
Presence of weak inter-cite couplings should not drastically destroy the integrability.
One may expect that weak couplings should lead to the onset of exponential soliton tales.
Thus, we can anticipate that formation of stable solitonic solutions should require
interaction energy to be large compared to the kinetic and 
Rabi energies.
This suggest that such solitons should be well spatially separated (in order to reduce the kinetic energy) and consist of only 
one species (to reduce the Rabi energy).
Results of \cite{JRLR14} indicate that solitons of this kind can arise spontaneously, provided
wavepacket spreading is weak.

It is informative to consider the case of $J=0$ corresponding to a very deep lattice
where neighbouring sites are decoupled.
In this case Eqs.~(\ref{Hsys}) can be transformed to the following form:
\begin{equation}
 \frac{dz_n}{dt} = -\Omega\sqrt{1-z_n^2}\sin\chi_n,\quad
 \frac{d\chi_n}{dt} = g\rho_n z_n + \frac{\Omega z_n}{\sqrt{1-z^2}}\cos\chi_n,
\label{zsys}
 \end{equation}
where
\begin{equation}
 z_n = \frac{A_n - B_n}{\rho_n},\quad
 \chi_n = \alpha_n - \beta_n.
\end{equation}
Equations (\ref{zsys}) originate from the Hamiltonian
\begin{equation}
 \tilde H_n(z_n,\,\chi_n) = 
 \frac{g\rho_nz_n^2}{2} - \Omega\sqrt{1-z_n^2}\cos\chi_n .
 \label{Htilde}
\end{equation}
Qualitative behavior 
described by the Hamiltonian (\ref{Htilde}) crucially depends on the parameter 
\begin{equation}
 \Lambda_n = \frac{g\rho_n}{\Omega}.
\end{equation}
If $\Lambda_n<1$, 
any trajectory of (\ref{zsys}) is rotation in phase space around one of the center fixed points 
located at $z_n=0$, $\chi_n=k\pi$, $k=0,1$ \cite{Oberthaler1}.
It corresponds to oscillations of the population imbalance $z_n$ with zero mean.
If $\Lambda_n>1$, the fixed point located at $z_n=0$, $\chi_n=\pi$  undergoes the pitchfork
bifurcation and transforms into three fixed points, one saddle and two centers.
The newborn center fixed points are located at $z_n=\pm\sqrt{1-\Lambda_n^{-2}}$, $\chi_n=\pi$.
Each of them is surrounded by a domain of population imbalance
oscillations with non-zero mean, that is,
one species dominates over another.
This regime can be referred to as the internal self-trapping.
Further growth of the interaction strength $g$ moves the center fixed points towards the limiting values
of population imbalance, $z_n=\pm 1$, and phase space area corresponding to the internal self-trapping increases.

\section{Wavepacket dynamics for various initial conditions: main features}
\label{Undistort}

\begin{figure}[h!tb]
\centerline{\includegraphics[width=0.8\textwidth]{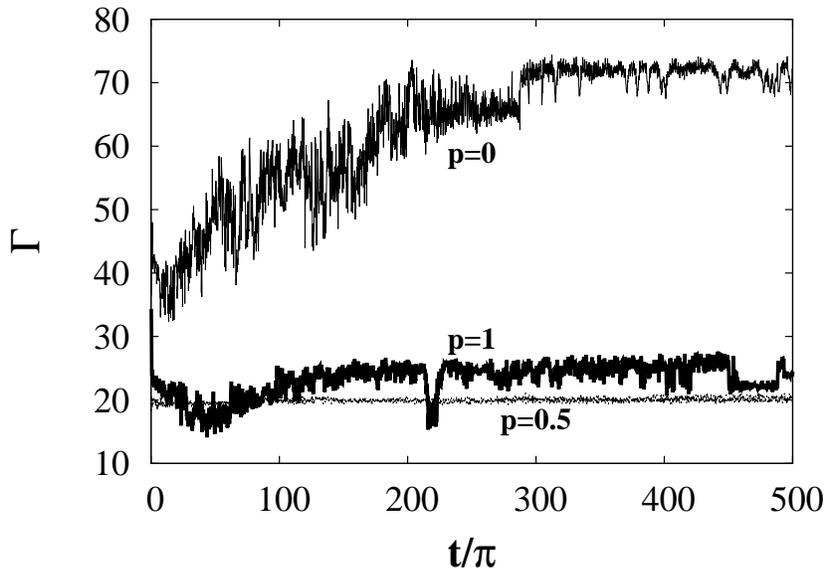}}
\caption{Participation ratio vs time for various forms of initial conditions.}
\label{fig-gamma}
\end{figure}

In the present work
we are aimed in studying formation of immiscible (i.e., one-species)
solitons. Therefore we
consider the regime of relatively strong nonlinearity $E_{\text{int}}/J\gg 1$,
where $E_{\text{int}}= H_{\text{int}}(t=0)$.
We use the following initial condition:
\begin{equation}
 a_n(t=0) = A(p)\exp\left[-\frac{n^2}{4\sigma^2}\right]\cos{\pi pn},\quad
 b_n=0,
\label{init}
\end{equation}
where  $-N\le n\le N$, 
the factor $A$ is determined by the normalization condition (\ref{norm}).
Parameter $\sigma$ characterizing wavepacket width is taken of 10.
According to results of \cite{JRLR14,TrSmerzi},
spatial dynamics of condensate depends on $E_{\text{int}}$ and 
the parameter $p$ describing spatial modulation.
We consider three values of $p$:
$p=0$ corresponding to the in-phase state,
$p=1$ corresponding to the state with checkboard sequence of phases at lattice sites,
and $p=0.5$ corresponding to the checkboard sequence, where all occupied sites are separated
from each other by unoccupied ones.

Throughout this paper we use the following values of parameters:
$J=2$, $\Omega=1$, $E_{\text{int}}=5$. 
It corresponds to the regime of strong nonlinearity anticipating
extensive creation of localized states due to self-trapping. 
As values of the tunneling rate $J$ and Rabi frequency $\Omega$ are of the same order,
one should expect strong interplay between spatial and internal dynamics.
In numerical simulation, we used lattices with $N$ ranging from 4000 to 10000.
A particular value for each calculation is chosen 
in order to avoid the influence of boundaries, therefore, 
it depends on a rate of spatial wavepacket expansion.

To quantify self-trapping, it is reasonable to consider
participation ratio 
\begin{equation}
 \Gamma=\frac{1}{\sum\limits_{n} \rho_n^2}
\end{equation}
being approximate number of occupied lattice sites.
Figure \ref{fig-gamma} showing the time dependence of participation ratio confirms the significance of spatial modulation.
Notably, self-trapping observed in the case of checkboard sequences $p=0.5$ and $p=1$ 
is much stronger than in the case of the in-phase state $p=0$.
\begin{figure}[!htb]
\begin{center}
\includegraphics[width=0.8\textwidth]{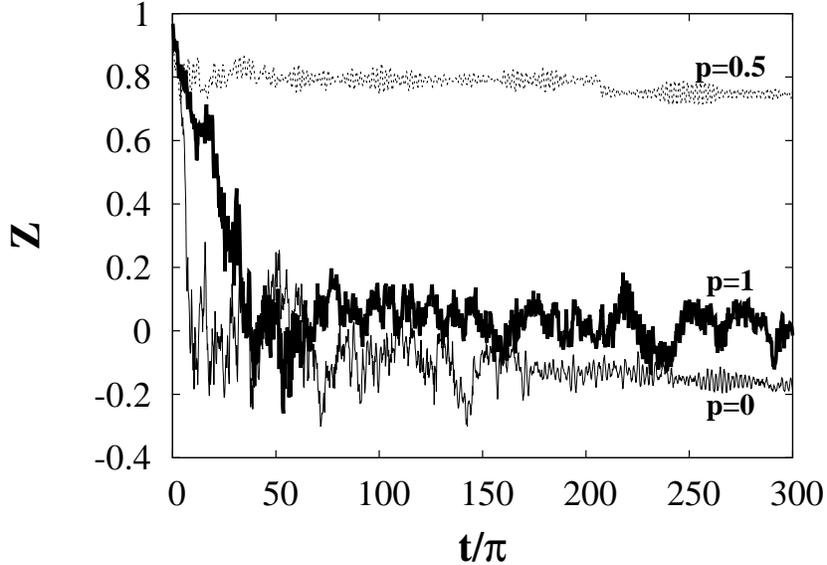}
\caption{Total population imbalance vs time for various forms of initial conditions.}
\label{fig-z}
\end{center}
\end{figure}
\begin{figure}[!phtb]
\begin{center}
\includegraphics[width=0.6\textwidth]{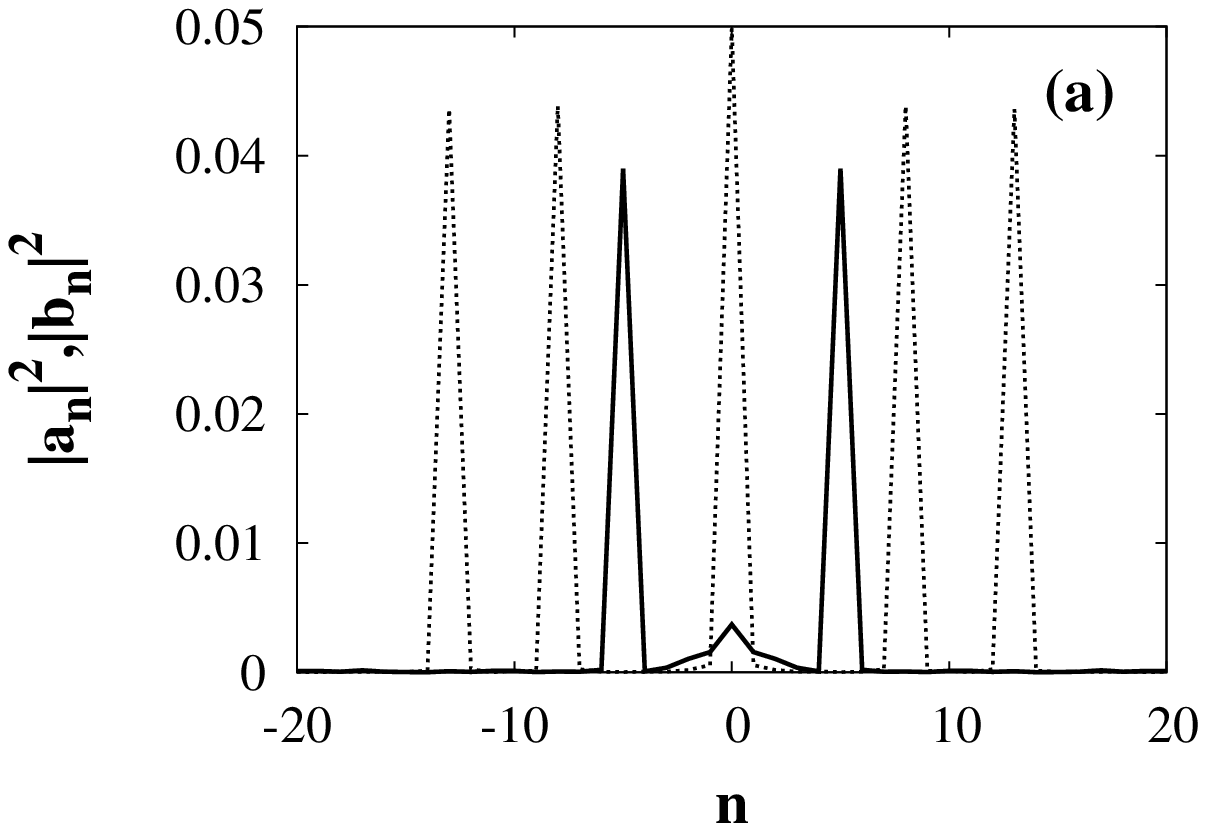}\\
\includegraphics[width=0.6\textwidth]{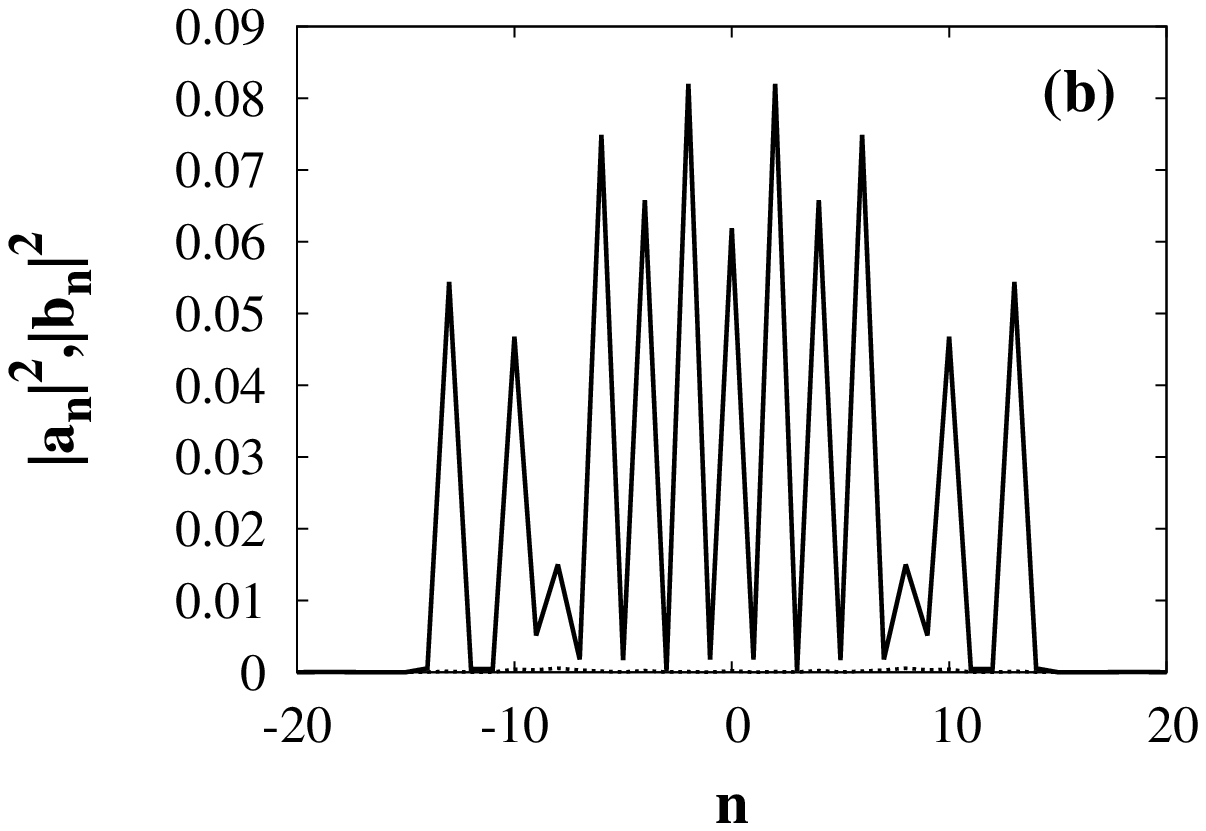}\\
\includegraphics[width=0.6\textwidth]{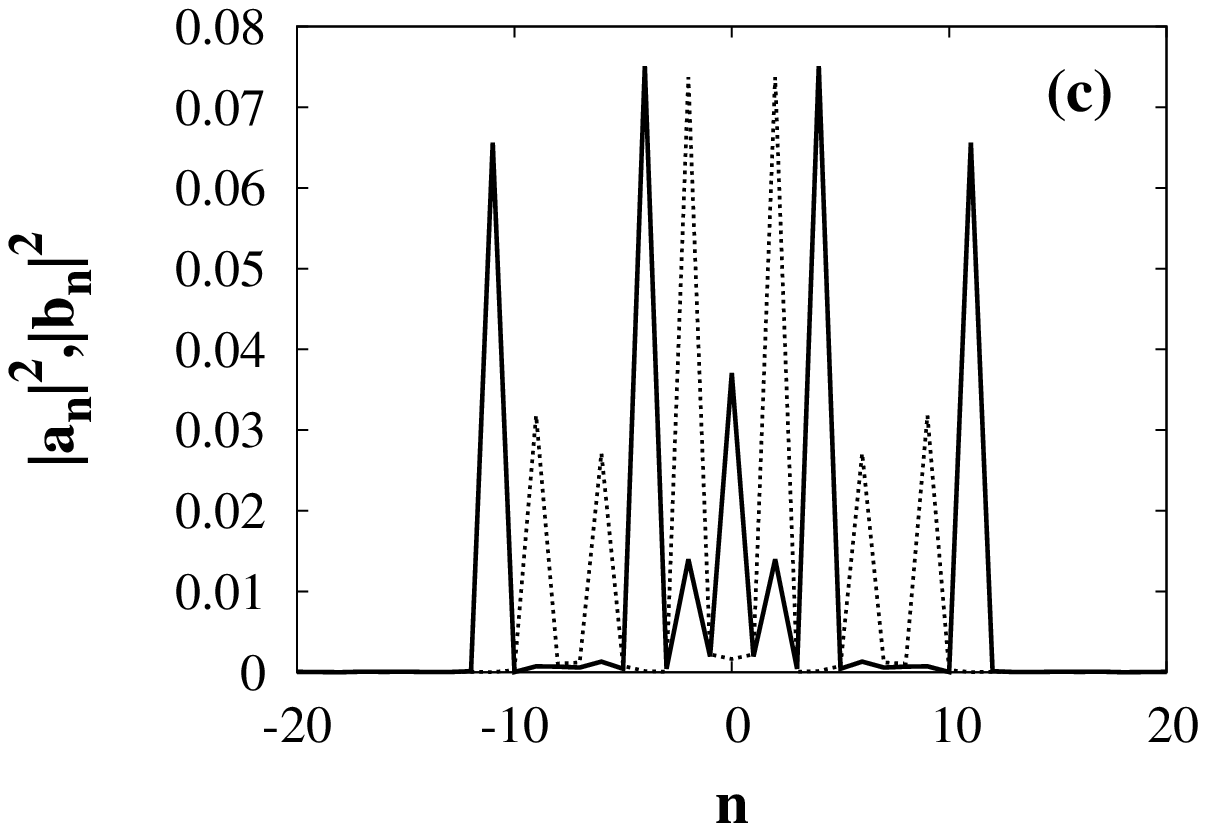}
\caption{Squared modulo of wavefunctions corresponding to the first (solid) and second (dotted) species at $t=500\pi$.
(a) $p=0$, (b) $p=0.5$, (c) $p=1$.}
\label{fig-field}
\end{center}
\end{figure}
\begin{figure*}[!htb]
\includegraphics[width=0.48\textwidth]{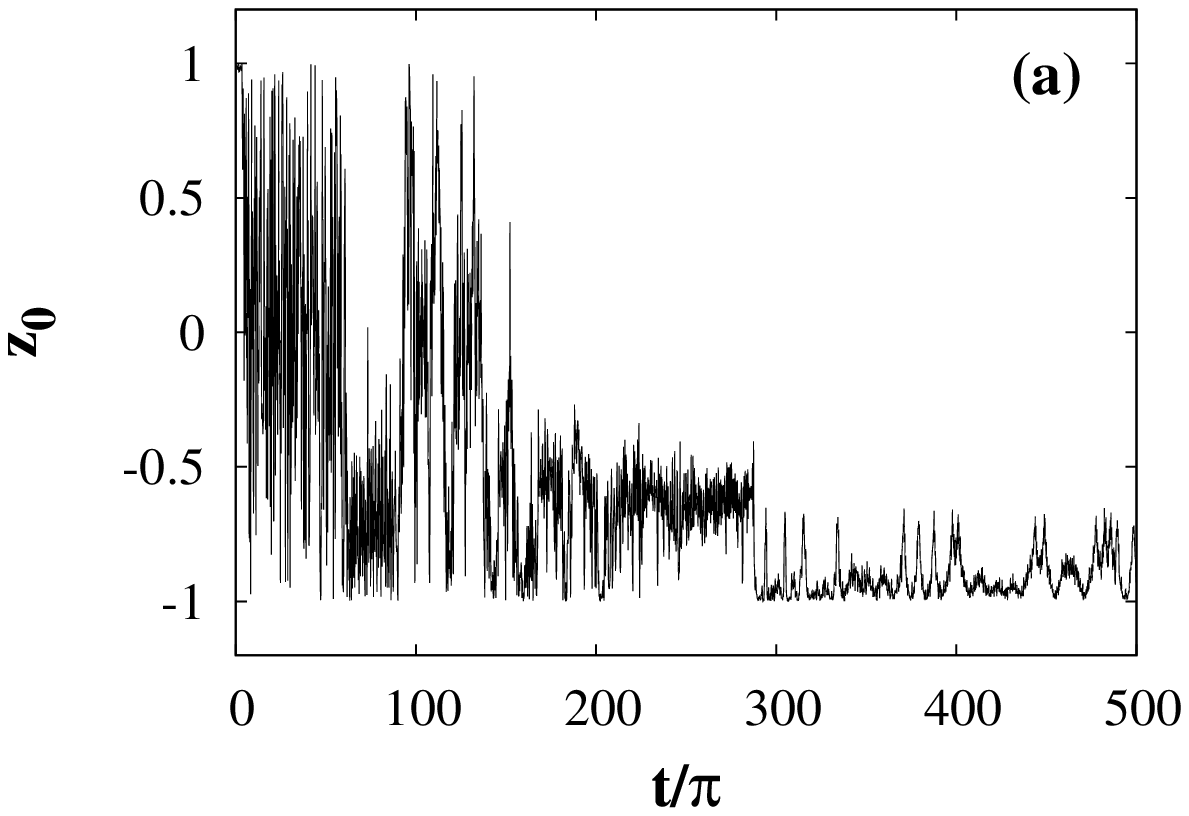}
\includegraphics[width=0.48\textwidth]{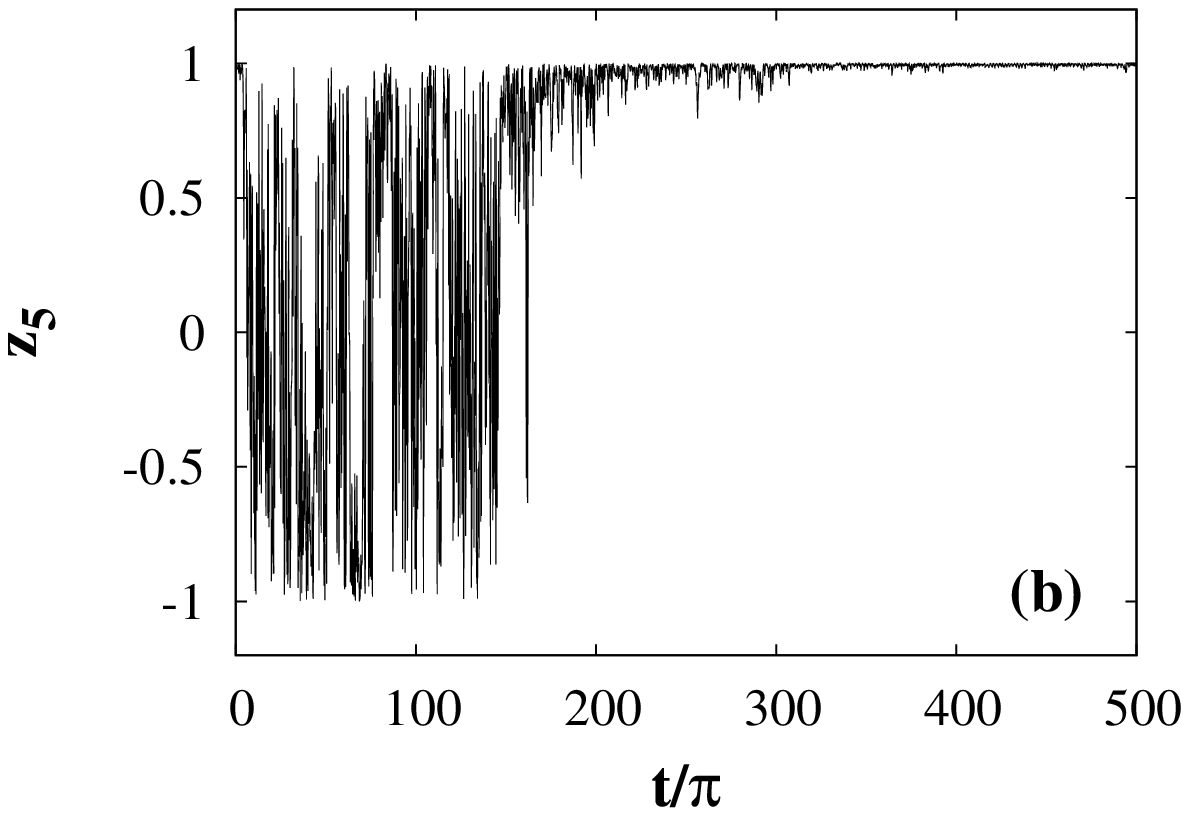}\\
\includegraphics[width=0.48\textwidth]{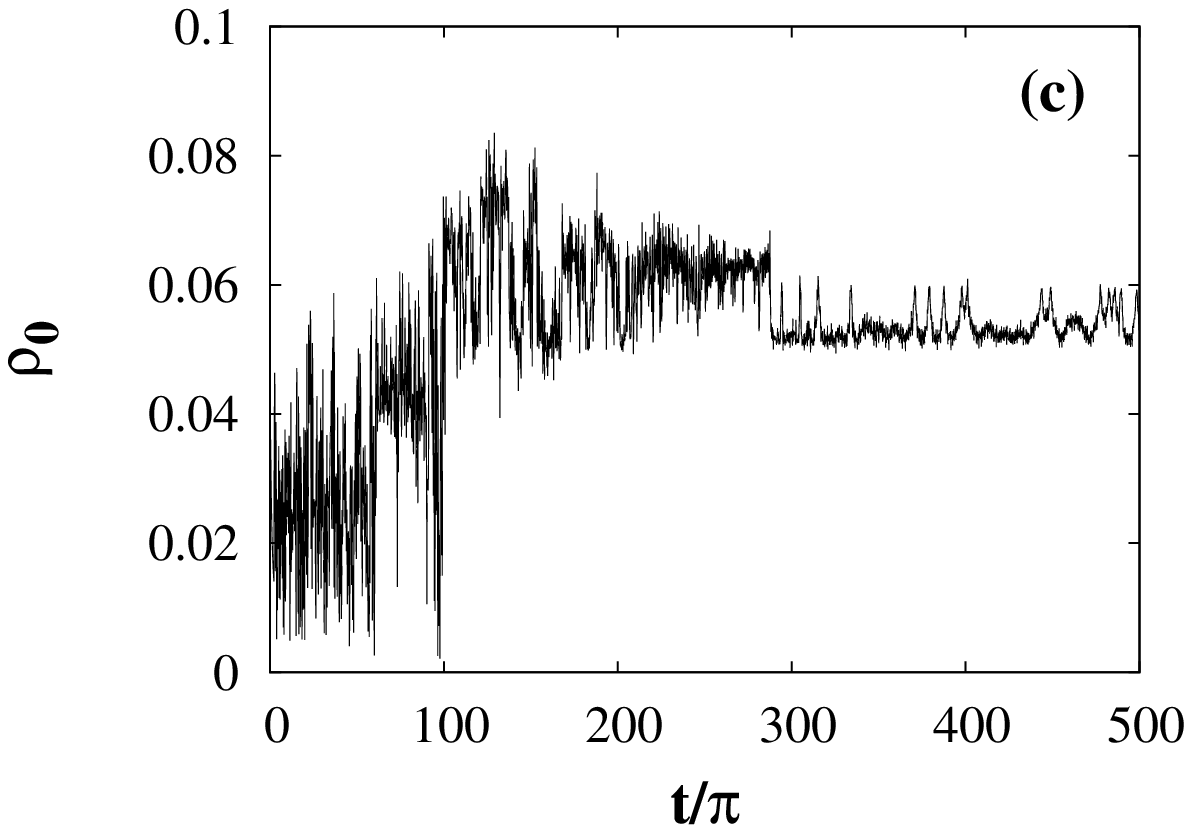}
\includegraphics[width=0.48\textwidth]{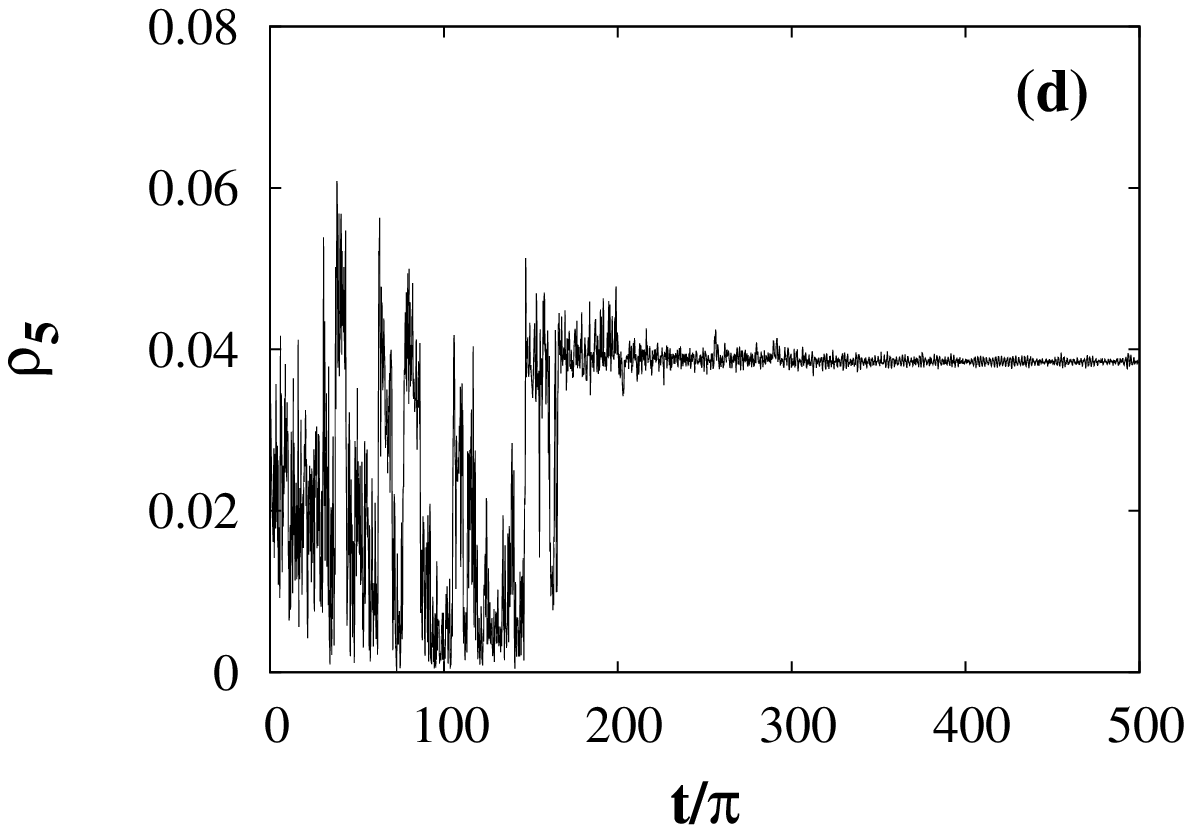}
\caption{Population imbalance (panels (a) and (b)) and population (panels (c) and (d)) at the lattice sites with $n=0$ ((a) and (c))
and $n=5$ ((b) and (d)).}
\label{fig-cell}
\end{figure*}

As it was shown in \cite{JRLR14},
spatial self-trapping is accompanied by internal self-trapping, i.e.,
Rabi inter-species oscillations are suppressed and
each soliton is predominantly formed by one species.
We can call such solitons as {\it immiscible} solitons.
As long as condensate fractions corresponding to the first and second species generally differ,
total population imbalance
\begin{equation}
Z=\sum\limits_n \rho_nz_n = \sum\limits_n |a_n|^2 - \sum\limits_n |b_n|^2.
 \label{Z}
\end{equation}
can significantly deviate from zero.
Time dependence of the total population imbalance is shown in Fig.~\ref{fig-z}. It is highly irregular
indicating
incoherence of Rabi oscillations running at different lattice sites.
The first species notably dominates in the regime of the strongest self-trapping, corresponding to $p=0.5$.
In contrast, the curves corresponding to $p=0$ and $p=1$ imply the presence of immiscible solitons corresponding to both species
in comparable amounts.
These suppositions are well confirmed by Fig.~\ref{fig-field} illustrating wave states at $t=500\pi$. 
Each of the panels (a)--(c) represents series of localized immiscible states.
In the case of $p=0.5$, all such states correspond to the first species,
while the whole pattern
looks like slightly distorted initial condition.
In contrast, both species are present in the cases of $p=0$
and $p=1$, and there is no apparent similarity 
between the wavepacket patterns and the corresponding initial conditions.
Absence of the similarity suggests that
the patterns depicted in Fig.~\ref{fig-field}(a) and (c) should result
from complicated evolution. 

Metamorphoses happening to wavepackets are revealed in 
time dependences of population imbalance
$z_n$ and population $\rho_n$ on individual lattice sites where solitons are pinned.
Let's consider the case of $p=0$ and the states localized at $n=0$ and $n=5$. 
The corresponding results are presented in Figure \ref{fig-cell}.
On the initial stage, population imbalance experiences several bursts of stochastic oscillations until
becomes localized near the limiting values: of -1 (for $n=0$) or 1 (for $n=5$).
Comparing the time dependences of $z_n$ and $\rho_n$, one can see that stabilization
of $z_n$ coincides with stabilization of $\rho_n$.
It implies that the stabilization happens when matter flux through a site 
ceases.
To achieve it, this site should have much larger population than on neighbouring sites,
as this provides the spatial self-trapping \cite{Gati_Oberthaler}.
It means that solitons are nucleated on specific density fluctuations. 
There is close analogy to emergence of solitons on impurities under conditions of spatiotemporal chaos 
\cite{Alexeeva_Barashenkov-PRL00}.

According to Fig.~\ref{fig-field},
the state at $n=0$ consists of both species. Presence of the second 
species reveals itself in sporadic variations of $z_0$ and $\rho_0$ which are absent in the case of $n=5$.
So, it turns out that admixture of another species
acts as a destabilizing factor for dynamics, 
that is, well-isolated immiscible states should be the most stable configuration.

\begin{figure*}[h!tb]
\begin{center}
\includegraphics[width=0.8\textwidth]{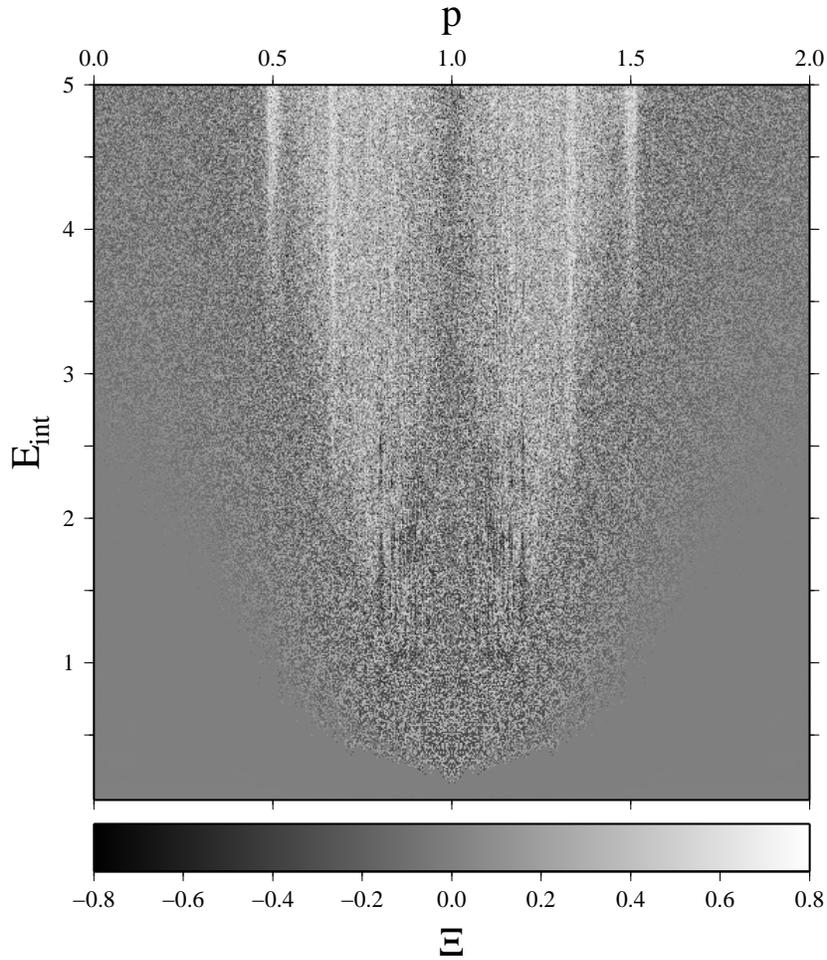}
\caption{The order parameter being 
the population imbalance averaged over time interval $[900\pi : 1000\pi]$ 
as function of parameters $E_\text{int}$ and $p$ characterizing the initial condition.} 
\label{fig-Zav}
\end{center}
\end{figure*}

To describe the resulting solitonic patterns,
we need a properly defined order parameter.
Total population imbalance looks as a good candidate
as it becomes non-zero as solitons appear, except for the case
when solitons of different species give equal contributions and therefore
compensate each other. In fact, the latter phenomenon can be readily considered as a rare event
and therefore neglected.
To eliminate contribution of non-solitonic states, we can average 
population imbalance over sufficiently long time interval.
This interval must correspond to large enough times when all transient regimes are completed,
and a wave field achieves some equillibrium state.
Thus, we arrive to the following definition:
\begin{equation}
 \Xi =\frac{1}{\Delta} \int\limits_{t'-\Delta}^{t'}Z(t)\,dt,
\end{equation}
where $t'=1000\pi$, $\Delta = 100\pi$. 
Figure \ref{fig-Zav} demonstrates phase diagram of condensate in coordinates $p$ and $E_{\text{int}}$.
Fine-grained structure of the diagram indicates high sensitivity of the order parameter to small changes
in initial conditions, that is typical for chaos. Nevertheless, 
one can see vertical stripes with the prevailence of white, in particular, at  
 $p=\pm 0.5$. These regions correspond to initial conditions which are more
favorable for the solitons of the first species. 
Uniform region below the fine-grained pattern corresponds to dynamics without self-trapping, when population imbalance
oscillates with zero mean.

\section{Monte-Carlo sampling}
\label{Monte-Carlo}

\subsection{Perturbation of initial conditions}

\begin{figure}[!htb]
\begin{center}
\includegraphics[width=0.48\textwidth]{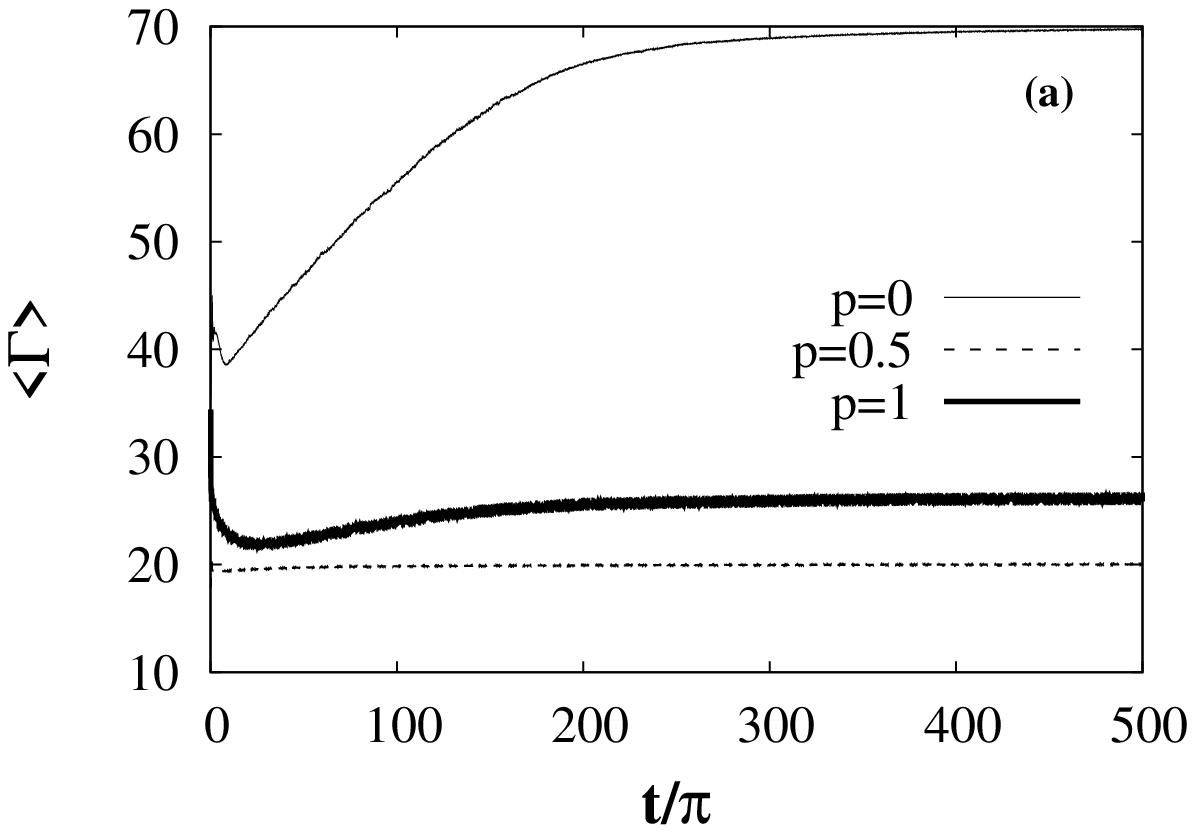}
\includegraphics[width=0.48\textwidth]{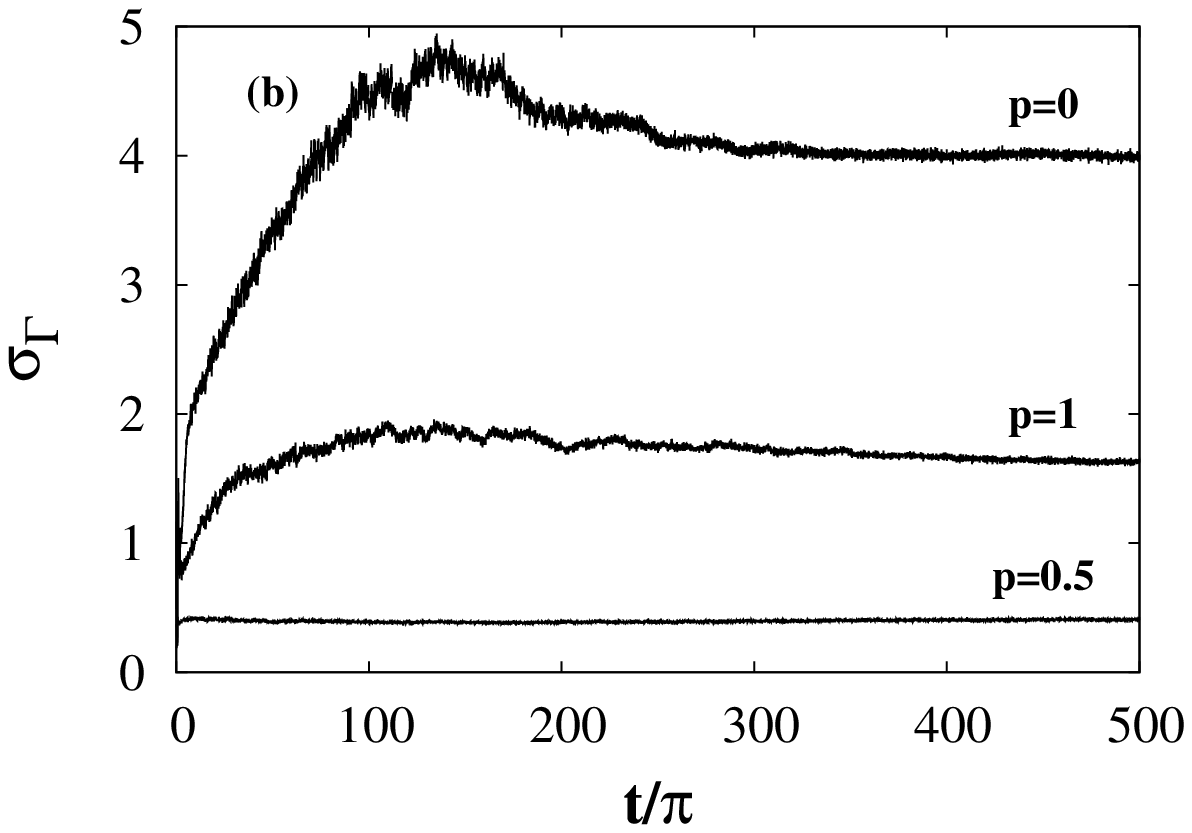}
\caption{Ensemble-averaged participation ratio (a) and its standard deviation (b) vs time.}
\label{fig-tr}
\end{center}
\end{figure}

Extreme sensitivity of wavepacket 
dynamics to small changes of initial conditions burdens analysis of physical condensate properties.
Therefore, it is reasonable to use some statistical averaging in order to smooth that sensitivity out.
Following this aim,
we impose random perturbation on an initial state
\begin{equation}
\begin{aligned}
 \mathbf{a}(t=0) &= \mathbf{a^{(0)}}(t=0) + \mathbf{\nu},\\
 \mathbf{b}(t=0) &= \mathbf{b^{(0)}}(t=0) + \mathbf{\xi},
\end{aligned}
 \label{nuxi}
 \end{equation}
where $\mathbf{a^{(0)}}(t=0)$ and $\mathbf{b^{(0)}}(t=0)$
are given by (\ref{init}).
Then we carry out the Monte-Carlo sampling over $\nu$ and $\xi$ realizations.
Components of vectors $\mathbb{\nu}$ and $\mathbb{\xi}$ 
are Gaussian random variables whose moments
obey the  following formulae:
\begin{equation}
 \left<\nu_k^*\xi_k\right> = 0,\quad
 \left<\nu_k^*\nu_l\right> = \left<\xi_k^*\xi_l\right> = \delta_{kl}\frac{\varepsilon}{\sigma\sqrt{2\pi}}\exp\left[-\frac{k^2}{2\sigma^2}\right],
\label{prop}
 \end{equation}
where $\delta_{kl}$ is the Kronecker symbol, and $\sigma$ takes on the same value as for original initial conditions.
Taking into account the normalization condition (\ref{norm}), we set $\varepsilon=0.01$ providing
weakness of the perturbation (\ref{nuxi}).

\begin{figure}[!htb]
\begin{center}
\includegraphics[width=0.48\textwidth]{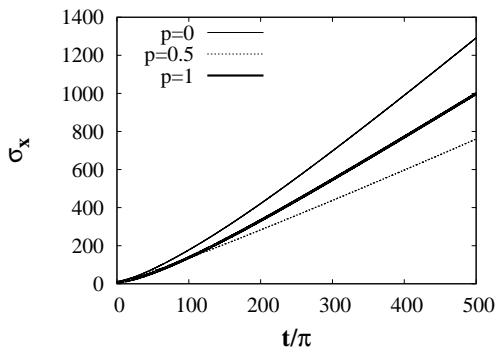}
\caption{Ensemble-averaged wavepacket width defined by formula (\ref{std}) as function of time.}
\label{fig-std}
\end{center}
\end{figure}

\begin{figure}[!htb]
\begin{center}
\includegraphics[width=0.48\textwidth]{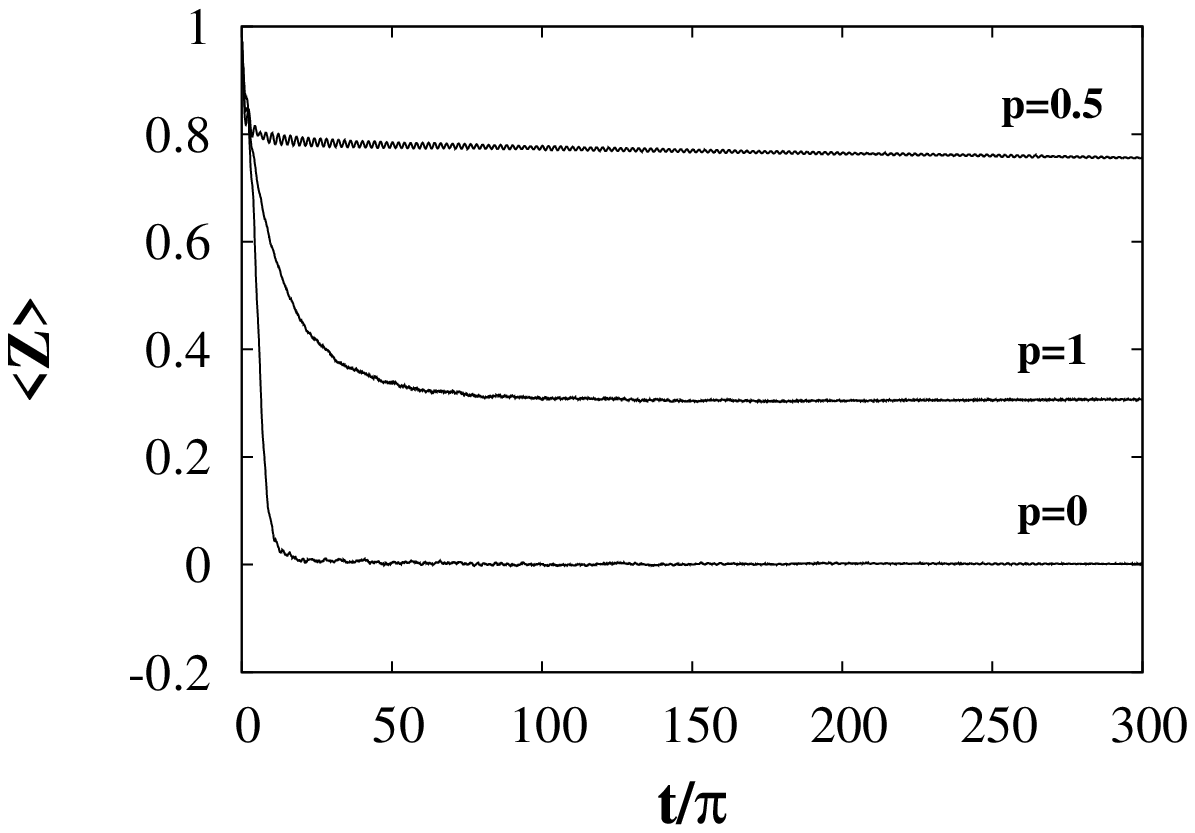}
\includegraphics[width=0.48\textwidth]{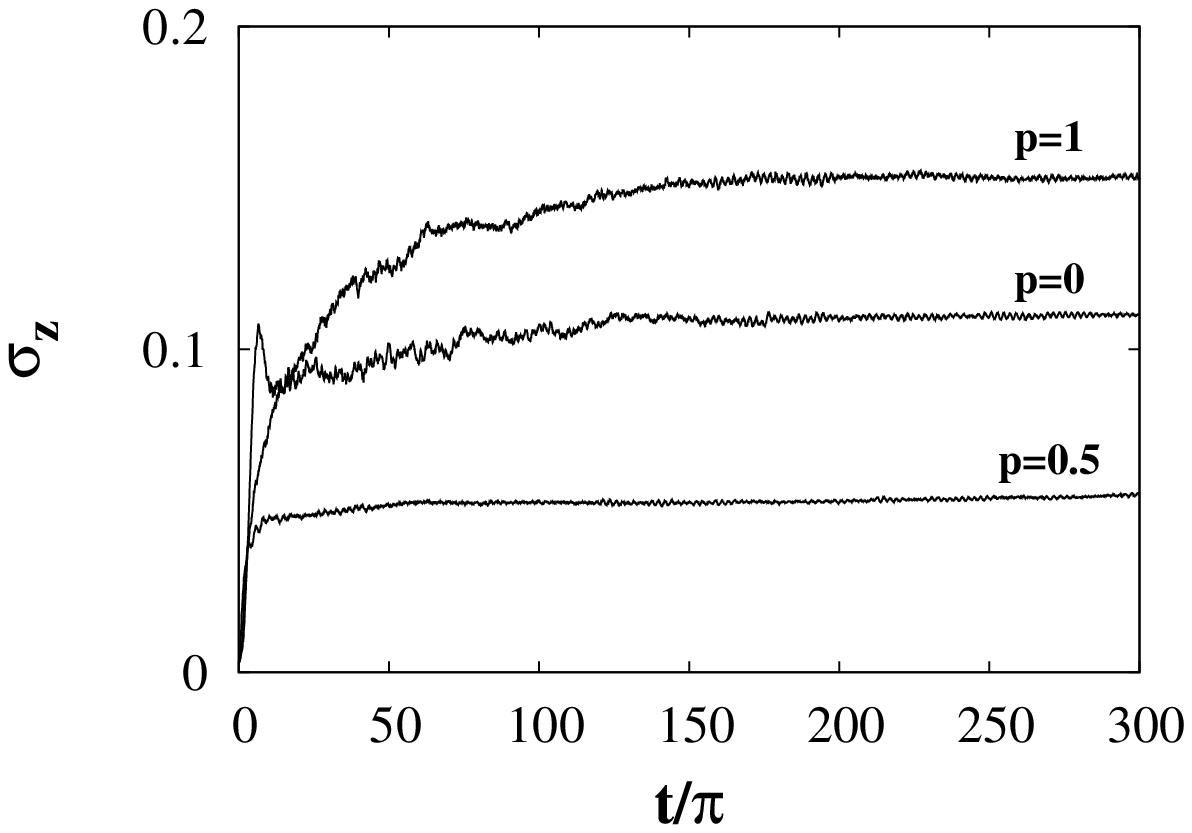}
\caption{Ensemble-averaged total population imbalance (a) and its standard deviation (b) vs time.}
\label{fig-z2}
\end{center}
\end{figure}

\subsection{Spatial vs internal dynamics}

Now let's consider ensemble-averaged picture of spatial and internal dynamics.
Ensemble-averaged participation ratio $\left<\Gamma\right>$ being a good indicator of the spatial self-trapping evolves 
in a very similar way as for non-distorted initial conditions (\ref{init}).
For all values of $p$, $\left<\Gamma\right>$ achieves a plateou and becomes almost constant, implying
onset of self-trapping (see Fig.~\ref{fig-tr}(a)).
In the case of the in-phase state $p=0$, the plateou corresponds to significantly larger values of $\left<\Gamma\right>$ reflecting
weaker self-trapping.
Comparing Figs.~\ref{fig-gamma} and \ref{fig-tr}(a),
one can conclude that
ensemble averaging smoothes out the small-scale fluctuations of participation ratio 
that are present in Fig.~\ref{fig-gamma}.
Time dependence of $\sigma_{\Gamma}$, being standard deviation of $\Gamma$, deserves an intent look.
In the case of $p=0.5$ it is almost constant, and the ratio $\sigma_{\Gamma}/\left<\Gamma\right>\simeq 0.02$
is very low revealing stable tendency to self-trapping.
In contrast, $\sigma_{\Gamma}$ for $p=0$ and $p=1$ varies non-monotonically with time. Its local maximum precedes 
stopping of wavepacket expansion and onset of the self-trapping.
After the self-trapping is happened, the standard deviation of $\Gamma$ becomes nearly constant,
and ratio $\sigma_{\Gamma}/\left<\Gamma\right>$ 
is approximately three times larger than in the case of $p=0.5$.
It indicates significantly higher diversity of solitonic configurations emerged.

Spatial expansion of a wavepacket can be described by the quantity
\begin{equation}
 \sigma_x = \frac{1}{\sqrt{M}}\sqrt{\sum_k \Delta_m^2}, \quad
 \Delta_m = \sum_n n^2\rho_n^{(m)} - \left(\sum_n n\rho_n^{(m)}\right)^2,
 \label{std}
\end{equation}
where index $m$ numbers realizations of perturbations $\mathbf{\nu}$ and $\mathbf{\xi}$, $M$ is total number of realizations.
Figure \ref{fig-std} shows that wavepackets are spreading ballistically despite of 
the spatial self-trapping and onset of solitons. It means coexistence of ballistic and localized
parts of the condensate. 
Comparing the data corresponding to different $p$, we see that the most fast spreading is observed for the in-phase state.

Time dependence of the ensemble-averaged total population imbalance $\left<Z\right>$
has remarkable differences with that for non-distorted initial conditions.
After rapid decreasing in the beginning, all the curves achieve a plateou
whose position depends on initial state (see Fig.~\ref{fig-z2}(a)).
In the case of $p=0.5$ the plateou corresponds to $\left<Z\right>=0.74$ reflecting dominance of the first species.
On the other hand, both species have the same statistical weights in the case of the in-phase state $p=0$.
Dynamics of the state $p=1$ corresponds to an intermediate regime, when dominance of the first species
is not so pronounced as in the case of $p=0.5$. Also, the case of $p=1$ demonstrates the largest values
of standard deviation $\sigma_Z$ indicating the highest sensitivity of Rabi oscillations
to small variations of initial conditions.

\subsection{Lyapunov analysis}
\label{Lyapunov}

As it was shown in the preceding section, dynamics of binary BEC in an optical lattice
in the regime of strong nonlinearity can exhibit highly irregular behavior.
Extreme sensitivity to small distortions of a wavepacket infers dynamical chaos.
As Eqs.~(\ref{TB2}) are nonlinear ODE, onset of chaos is not surprising \cite{ABDULLAEV1989,Hennig_Tsironis,Kolovsky-PRA09}.
As it was shown in \cite{Burgdorfer-depletion},
onset of dynamical chaos in BEC dynamics is closely related to the process of condensate depletion
due to partial thermalization.

Chaos strength can be quantified by means of the maximal Lyapunov exponent.
To determine it,
we can use a concept being the discretized version of the definition used in \cite{Burgdorfer-chaos}.
In that work one considers distance between two wave states in the Hilbert space
\begin{equation}
D(t) = \frac{1}{2}\left<\Psi'-\Psi\vert\Psi'-\Psi\right>,
\label{dist} 
\end{equation}
where $\Psi$ and $\Psi'$ are solutions of the nonlinear Schr\"odinger equation with infinitesimal difference 
in initial conditions.
Then the Lyapunov exponent can be determined as 
\begin{equation}
 \lambda = \lim\limits_{t\to\infty}\,\lim\limits_{D(0)\to 0}
 \frac{1}{2t}\ln\frac{D(t)}{D(0)}.
\end{equation}
In analogy with \cite{Burgdorfer-chaos}, we can define the Lyapunov exponent for the binary DNLSE
as
\begin{equation}
 \lambda = \lim\limits_{t\to\infty}\,\lim\limits_{d(0)\to 0}
 \frac{1}{2t}\ln\frac{d(t)}{d(0)},
 \label{lyap_gl}
\end{equation}
where
\begin{equation}
 d(t) = \frac{1}{2}\sum\limits_n (\delta a_n^*\delta a_n +
 \delta b_n^*\delta b_n ),
\end{equation}
\begin{equation}
\delta a_n =\tilde a_n - a_n,\quad
\delta b_n =\tilde b_n - b_n,
\end{equation}
and the tilde denotes a solution with slightly perturbed initial conditions.

Despite of the importance of $\lambda$ as a chaos descriptor,
it corresponds to the infinite-time limit and therefore doesn't give physically relevant 
information about intermittency in course of evolution \cite{Datta_Ramaswamy}.
Therefore, it is reasonable to consider finite-time Lyapunov exponent (FTLE)
\begin{equation}
 \lambda_{\Delta t}(\bar t) = \frac{1}{2\Delta t}\lim\limits_{d(t_0)\to 0}\ln\frac{d(t_0+\Delta t)}{d(t_0)},\quad
 \bar t = t_0 + \frac{\Delta t}{2}.
\label{FTLE}
 \end{equation}
\begin{figure}[!htb]
\begin{center}
\includegraphics[width=0.48\textwidth]{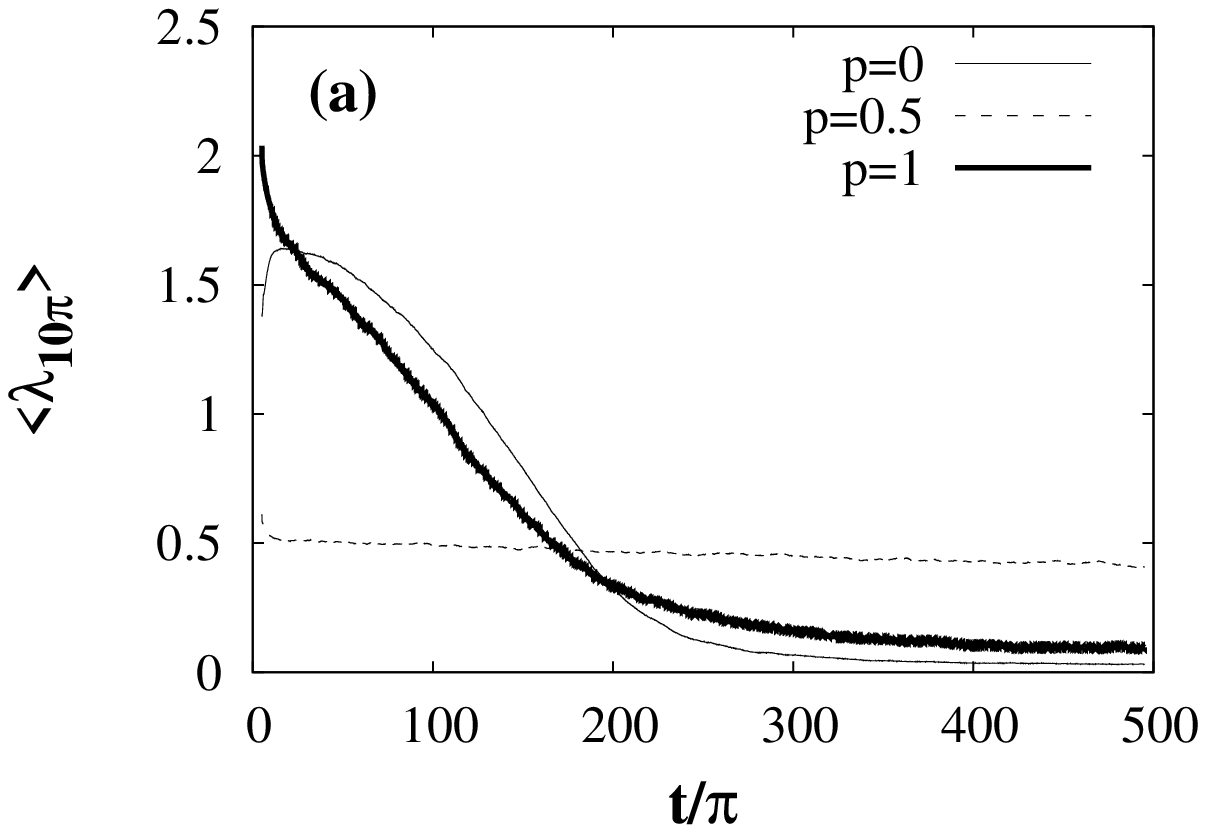}
\includegraphics[width=0.48\textwidth]{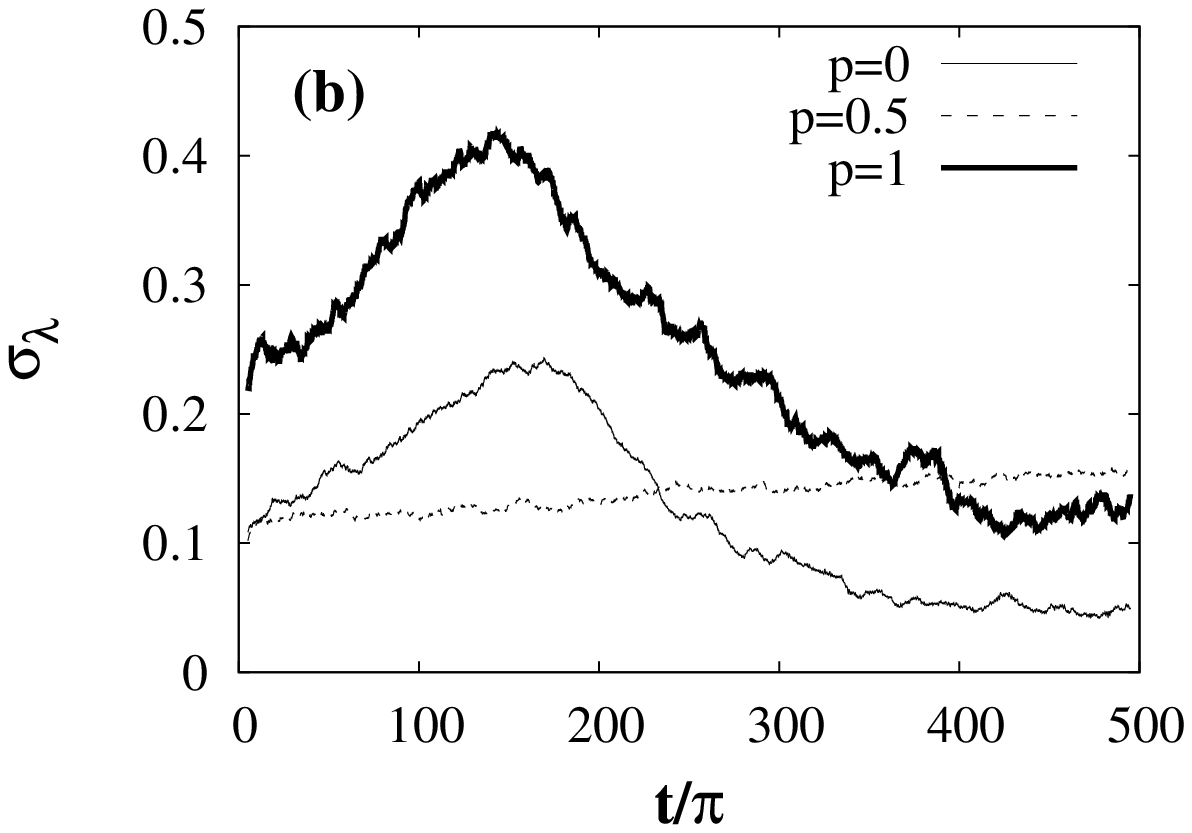}
\caption{Mean finite-time Lyapunov exponent with $\Delta t=10\pi$ (a) and its rms value (b) vs time.}
\label{fig-ftle}
\end{center}
\end{figure}

Figure \ref{fig-ftle}(a) demonstrates time dependence of the mean FTLE.
The curves corresponding to $p=0$ and $p=1$ reveal a crossover from chaotic to regular
dynamics with increasing time. 
Standard deviation of the FTLE varies non-monotonically with time, as is shown in Fig.~\ref{fig-ftle}(b).
Its maximum approximately corresponds to the time moment when nearly half of realizations
have come into the regular regime. 
Notably, the crossover happens almost simultaneously for $p=0$ and $p=1$.
As opposed to this, there is no apparent crossover in the case of $p=0.5$, and the mean FTLE, 
as well as its standard deviation, varies slowly with time.
Values of the mean FTLE correspond to relatively weak chaos as compared
to the initial stages in the cases of $p=0$ and $p=1$. Such behavior implies that
the wavepacket with $p=0.5$ is close to some long-living state and therefore doesn't experience
any drastic changes of dynamics. On the other hand, 
this case doesn't exhibit the self-stabilization as in the cases of $p=0$ and $p=1$.
It means the presence of weak but persistent instability, 
 that is, the initial state with $p=0.5$ should be rather regarded as
metastable one. 

\subsection{Energy analysis}
\label{Energy}

\begin{figure}[phtb]
\begin{center}
\includegraphics[width=0.46\textwidth]{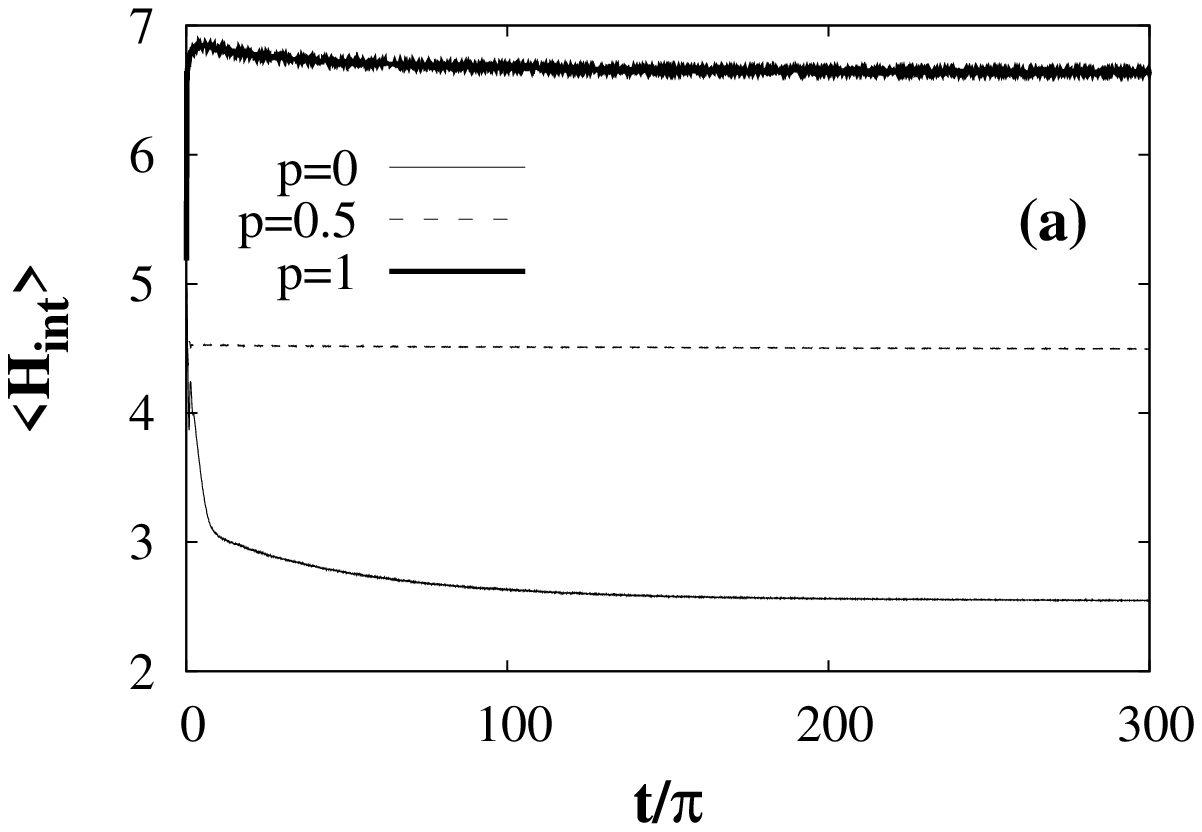}
\includegraphics[width=0.46\textwidth]{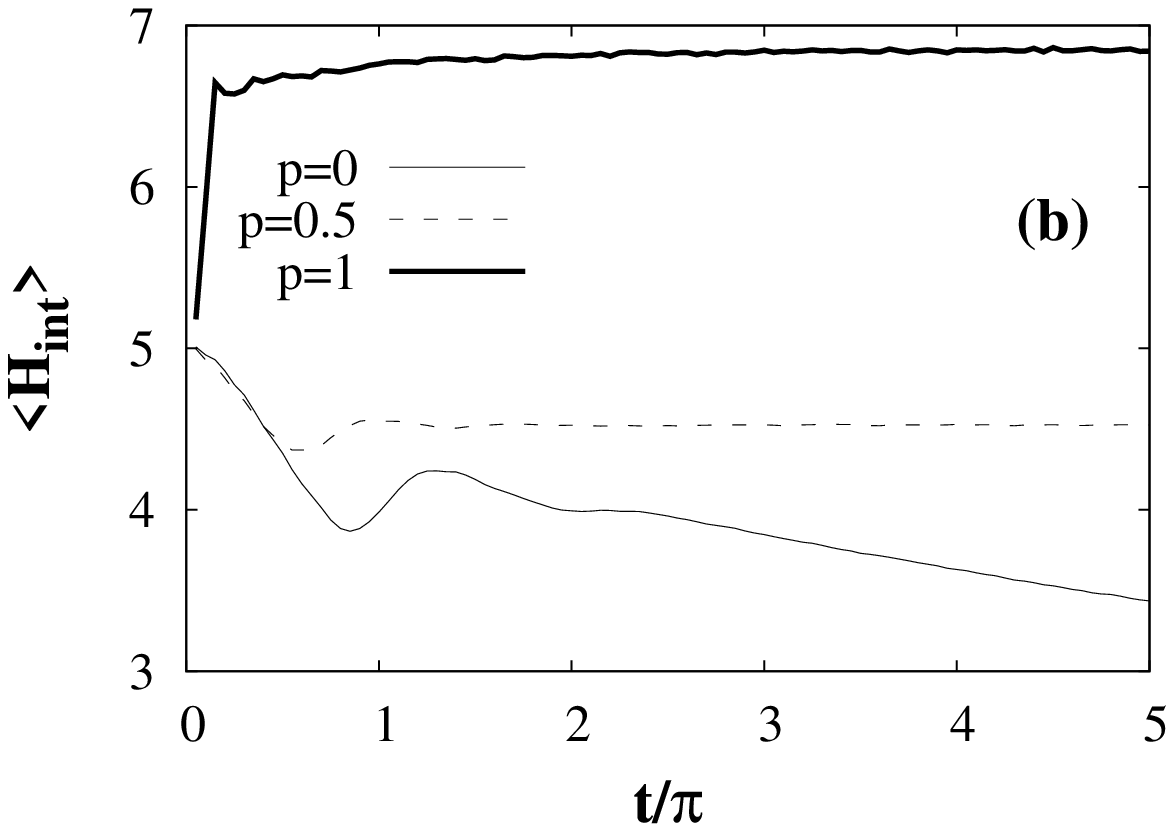}\\
\includegraphics[width=0.46\textwidth]{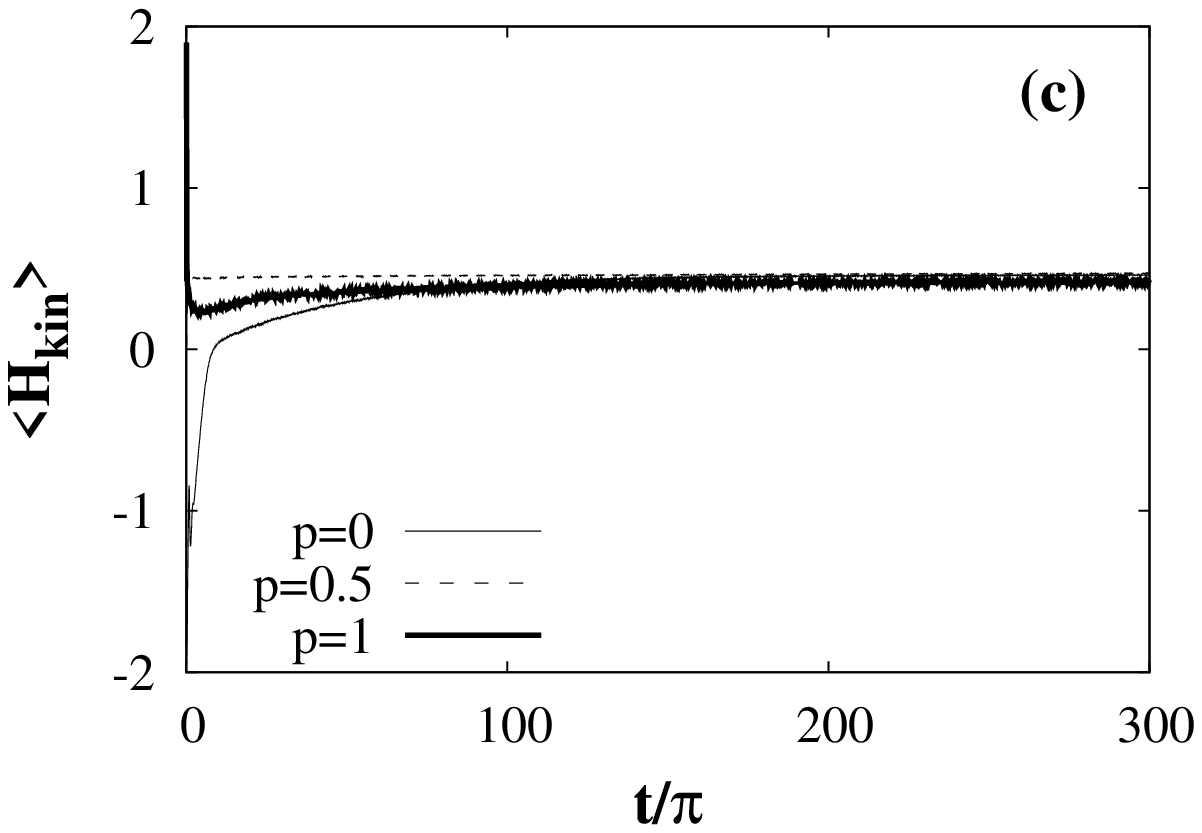}
\includegraphics[width=0.46\textwidth]{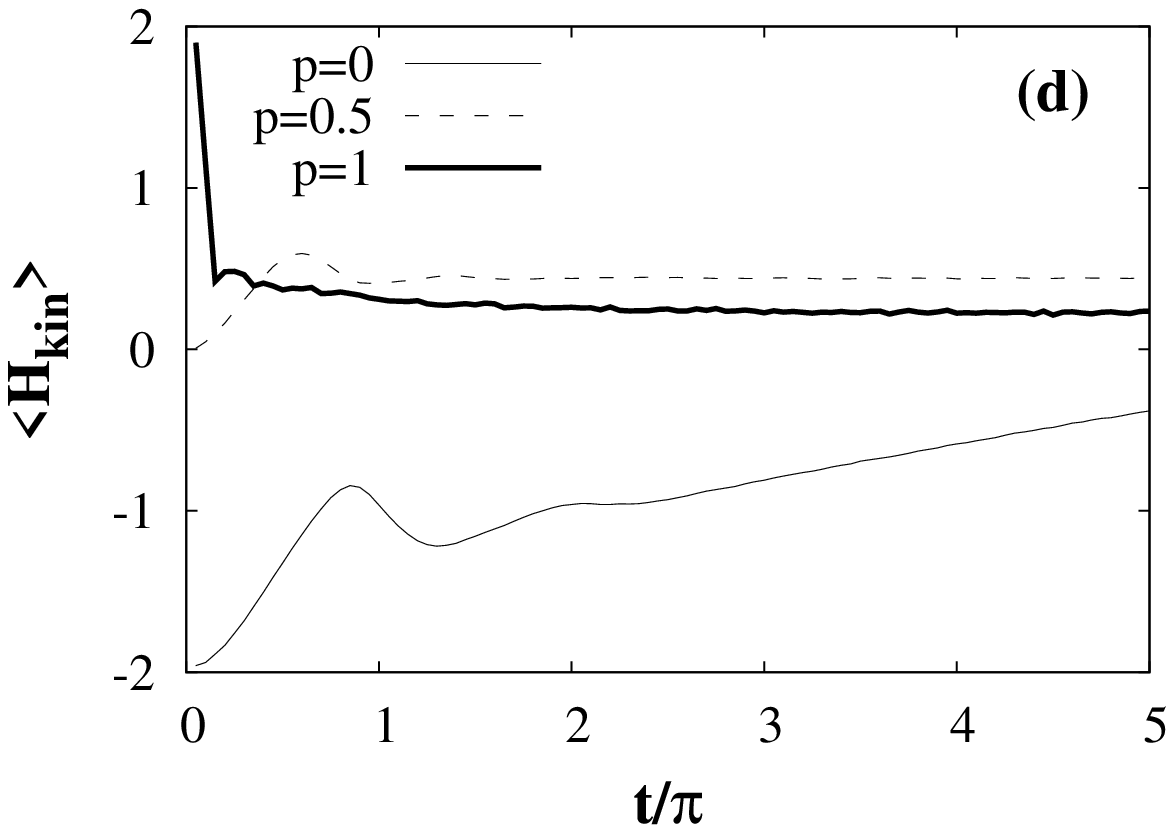}\\
\includegraphics[width=0.46\textwidth]{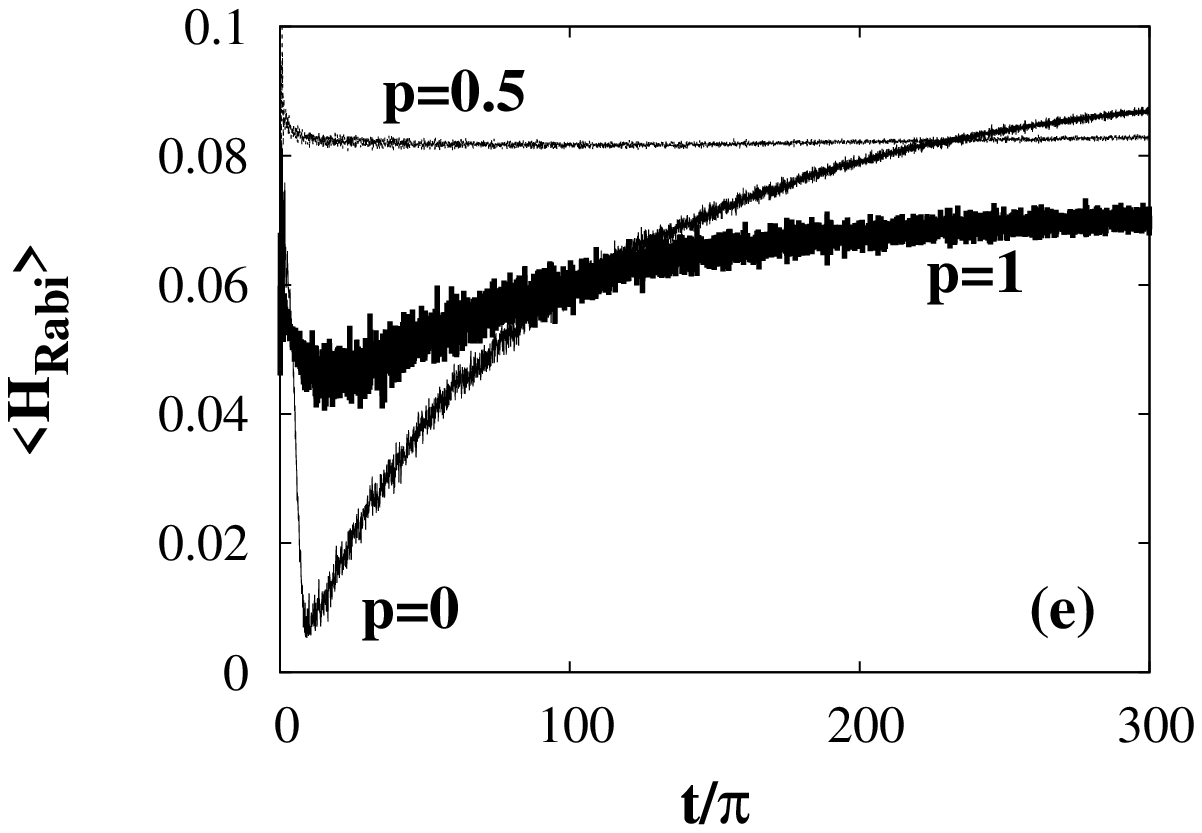}
\includegraphics[width=0.46\textwidth]{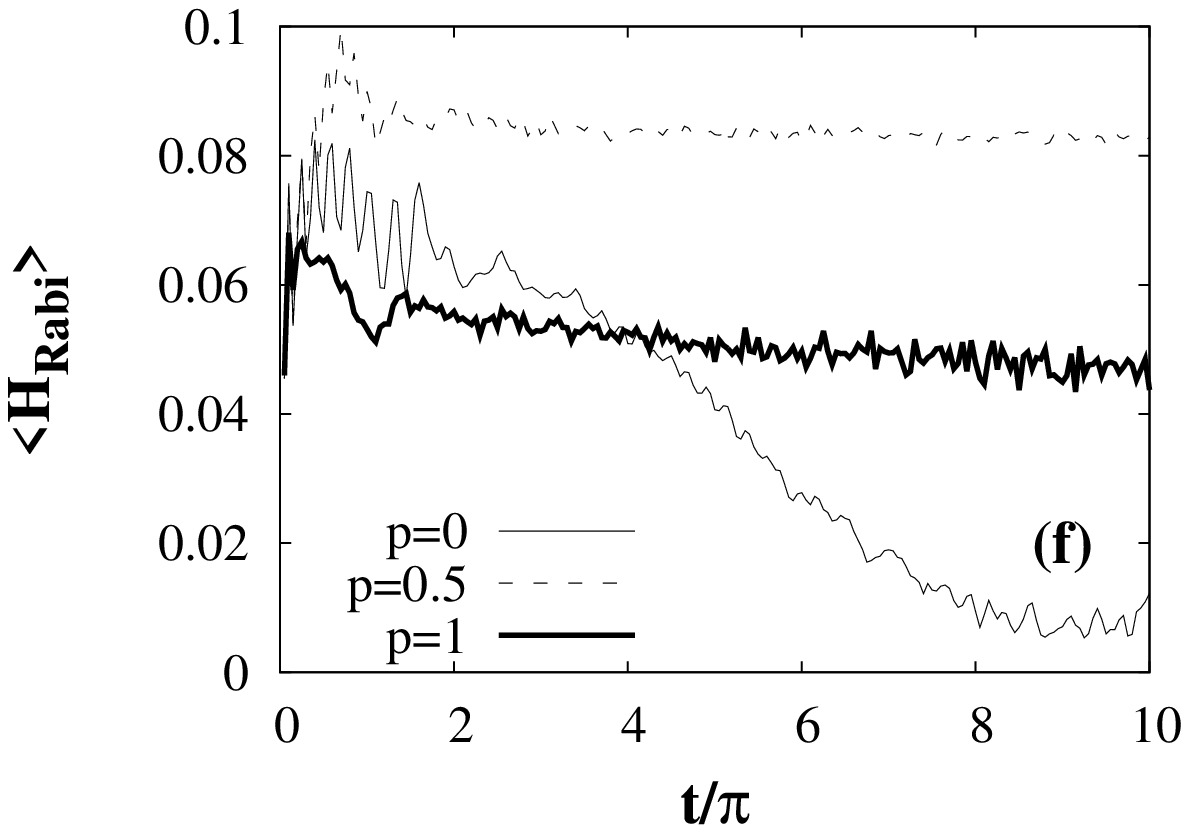}\\
\caption{Ensemble-averaged interaction (a-b), kinetic (c-d) and Rabi (e-f) energies vs time
for various initial conditions. Figures in the right column represent enlarged view of 
the beginning time intervals.}
\label{fig-en0}
\end{center}
\end{figure}

Energy conversion between components of the Hamiltonian (\ref{Ham0})
allows one to recover some specific features of wavepacket dynamics during the formation
of immiscible solitons.
Temporal variations of the interaction, kinetic and Rabi energies are presented 
in Fig.~\ref{fig-en0}.
One can see that exchange between the interaction and kinetic energies dominates,
while variations of the Rabi energy are much lower in magnitude.

Firstly, let's consider in detail the interaction energy (\ref{Eint}).
Owing to its definition, the interaction energy can serve a measure of wavepacket concentration
and soliton formation.
Quite surprisingly, the highest values of the interaction energy are observed in the case of $p=1$,
not in the case of $p=0.5$ where soliton formation is the most pronounced.
Another noticeable feature is extremely fast growth and decreasing of the interaction energy 
at the very beginning for $p=1$ and $p=0$, respectively.
Comparison of Figs.~\ref{fig-en0}(a) and (b) 
links the drastic variation of the interaction energy with the opposite variation (decreasing or increasing)
of the kinetic energy.
Intuitively, one can suggest that abrupt energy conversion should be a signature of strong instability 
and chaos ignition. So, it turns out that initial conditions with $p=0$ and 
$p=1$ are energetically unstable, in contrast to the case of $p=0.5$.

Another remarkable feature of the kinetic energy variations, presented in Fig.~\ref{fig-en0}(b),
is that the curves corresponding to different $p$ converge.
It is reasonable to suggest that the main contribution into the kinetic energy is given
from the ballistic fraction of condensate.
This ballistic fraction corresponds to small-amplitude
wavepackets emitted from the region occupied by solitons, i.e., the wavepacket origin.
However, one should keep in mind that volume of the ballistic fraction 
strongly depends on $p$. For instance, it is much smaller in the case of $p=0.5$ than
for other initial conditions.
This phenomenon can be explained by the difference in total energy determined by the sum (\ref{Ham0}).
Despite the absolute values of the kinetic energy are close to each other,
their fractions in the total energy are different. So, relative contribution of the kinetic term into 
the total energy is the largest for $p=0$ and smallest for $p=0.5$.

Rabi energy (\ref{ERabi}) can be regarded as a measure of inter-species miscibility.
Its time dependence is illustrated in Fig.~\ref{fig-en0}(c).
It is noticeable that the fast ignition of chaos for $p=0$ and $p=1$ in the beginning
is accompanied by abrupt fast variations of the Rabi energy. 
Firstly, the Rabi energy fastly grows due to generation of the second species,
and then there happens rapid spatial demixing of species.
In the case of $p=0$ the species become almost completely demixed, that is, the ensemble-averaged Rabi 
energy is nearly zero.
The demixing stage lasts several Rabi cycles. After that the Rabi energy starts increasing again,
indicating creation of miscible states in the ballistic fraction.
However, smallness of the Rabi energy variations infers smallness of population
of miscible states.
The main contribution into the miscible fraction is given from emitted ballistic
wavepackets. Indeed, they have significantly lower density than localized states and, therefore,
don't experience the internal self-trapping.

\begin{figure}[!htb]
\begin{center}
\includegraphics[width=0.48\textwidth]{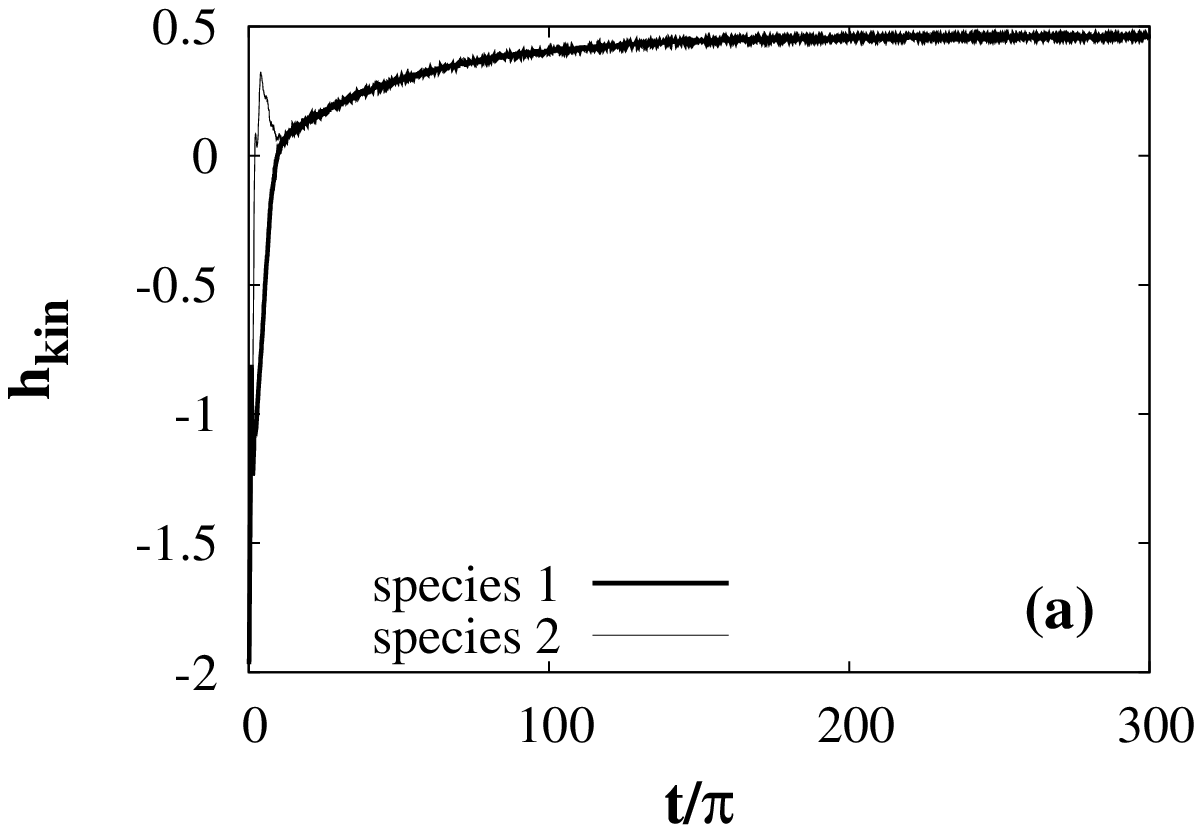}
\includegraphics[width=0.48\textwidth]{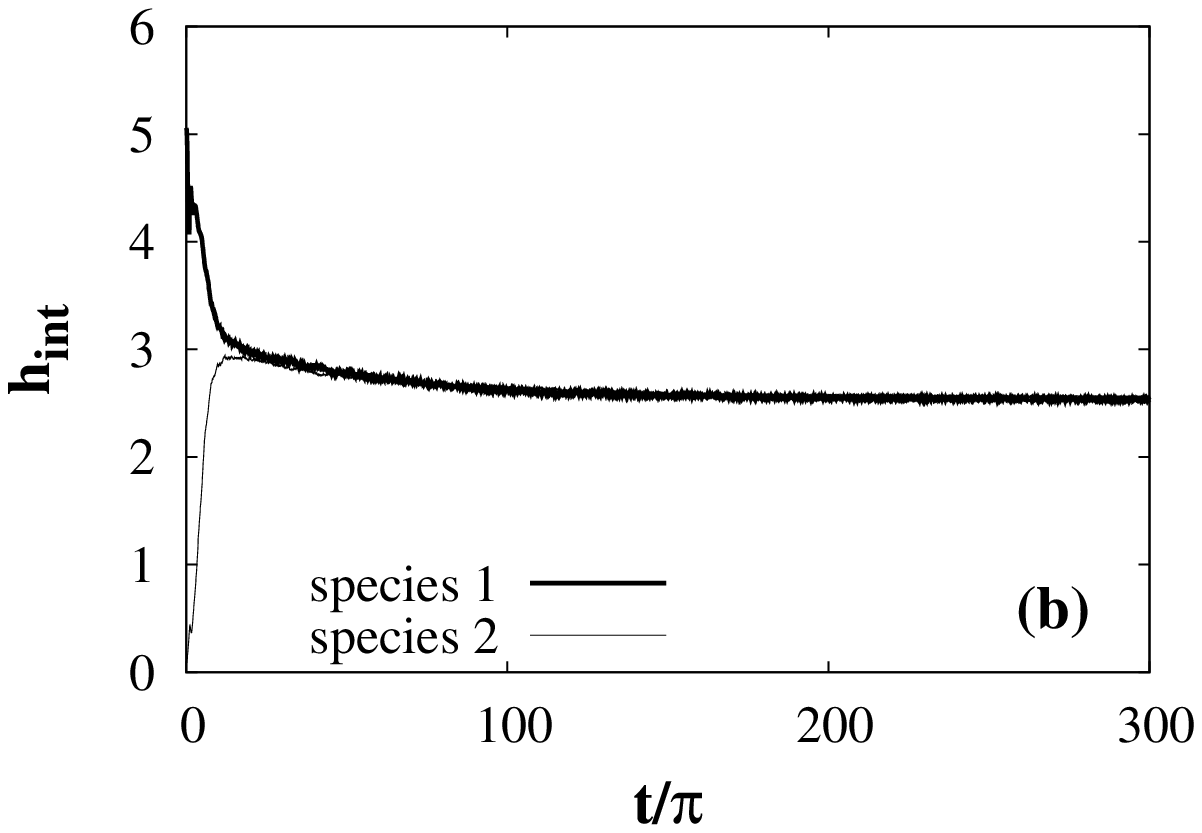}\\
\includegraphics[width=0.48\textwidth]{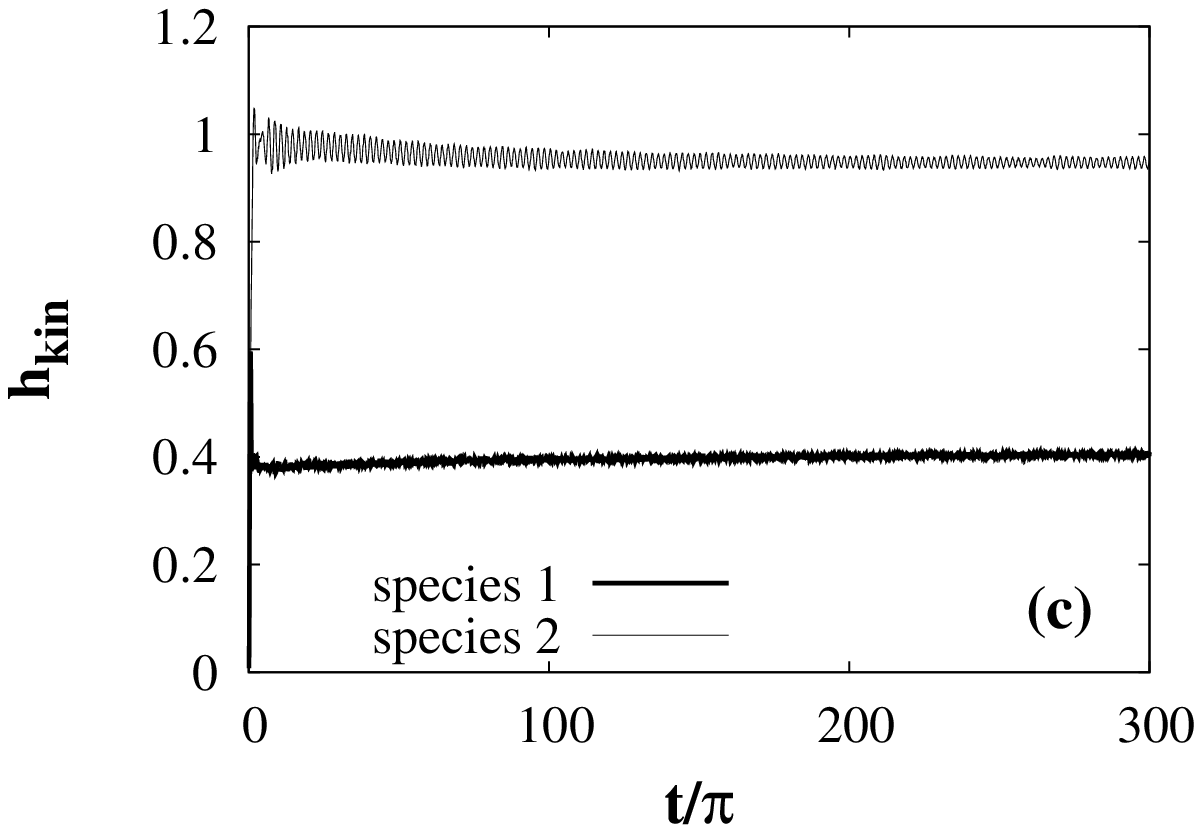}
\includegraphics[width=0.48\textwidth]{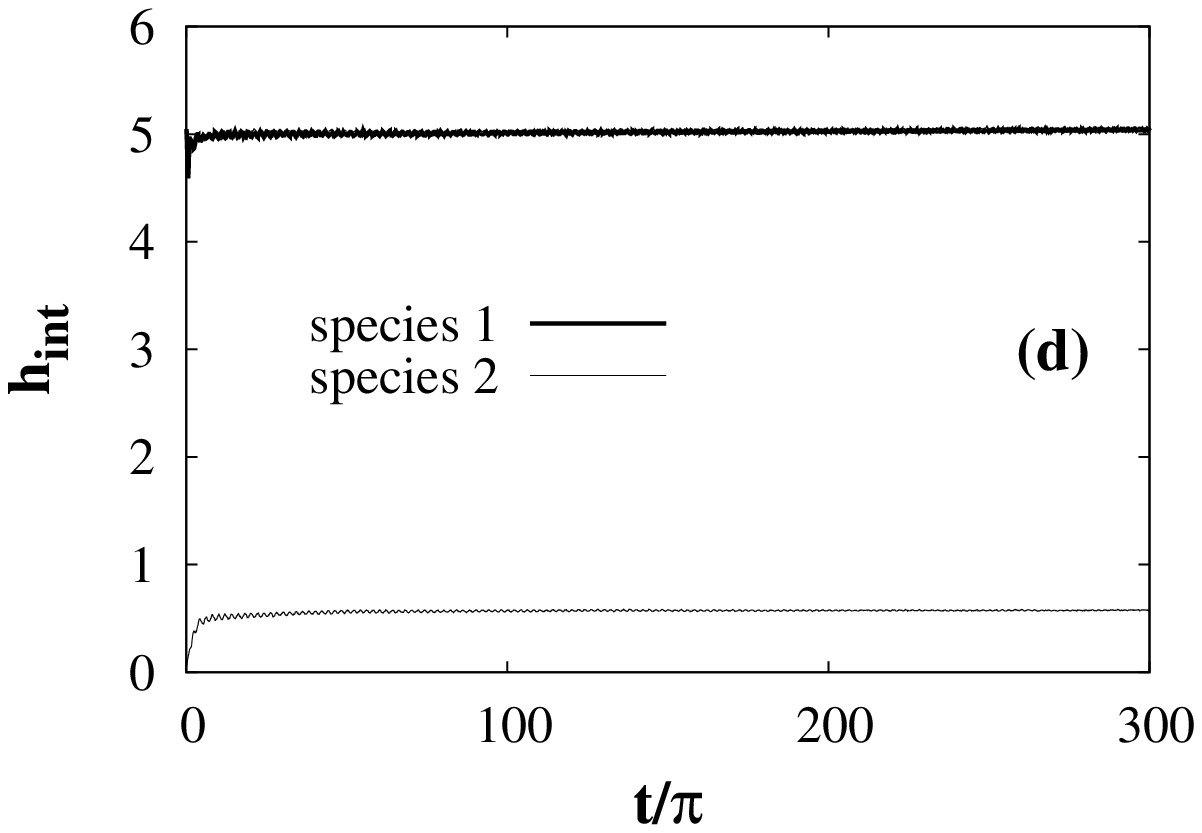}\\
\includegraphics[width=0.48\textwidth]{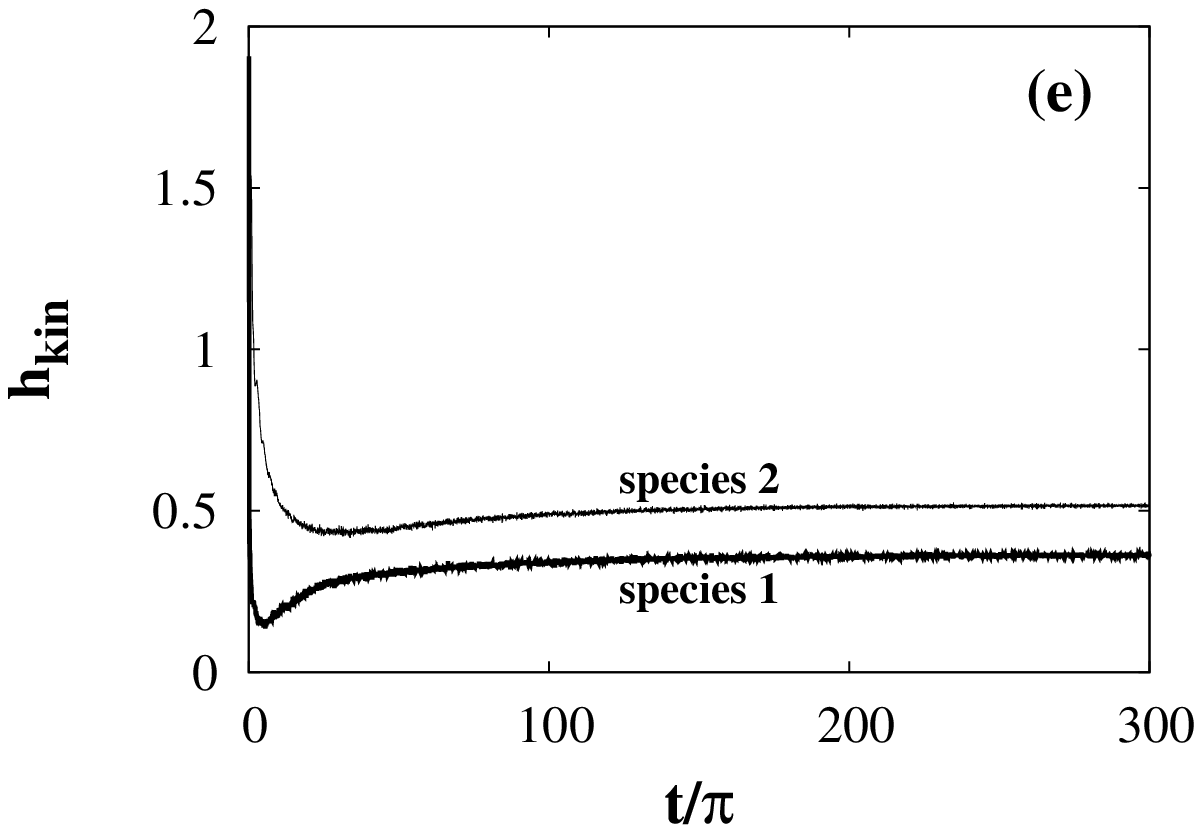}
\includegraphics[width=0.48\textwidth]{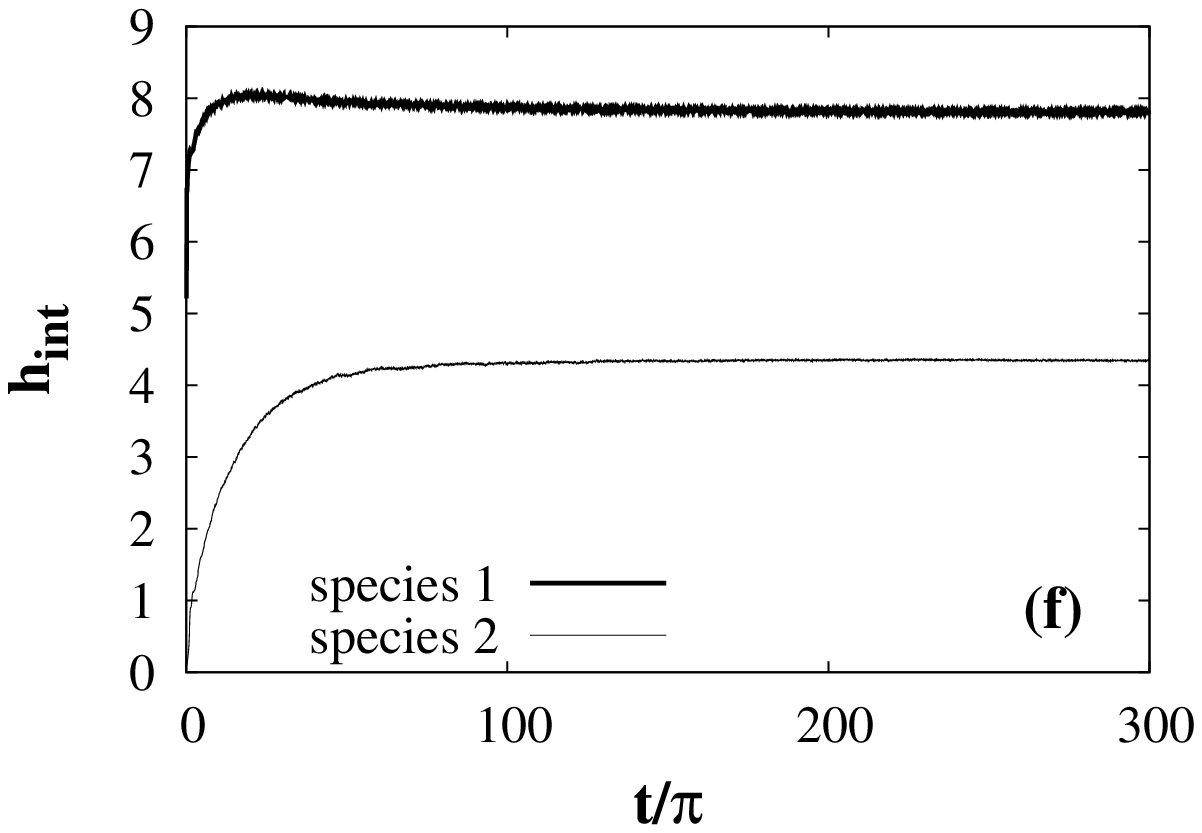}
\caption{One-species kinetic (a, c, e) and interaction (b, d, f) energy densities as functions of time for initial
states with $p=0$ (a, b), $p=0.5$ (c, d), $p=1$ (e, f).}
\label{fig-enr}
\end{center}
\end{figure}
%

It is informative to study how the kinetic and interaction energies are distributed
over species. It can be done by means of the quantities
\begin{equation}
 h_{\text{kin}}^{a,b} = \frac{\left<H_{\text{kin}}^{a,b}\right>}{\left<\rho_{a,b}\right>},\quad
 h_{\text{int}}^{a,b} = \frac{\left<H_{\text{int}}^{a,b}\right>}{\left<\rho_{a,b}\right>},
 \label{relative}
\end{equation}
where 
\begin{equation}
\begin{aligned}
H_{\text{kin}}^a &= -J\sum\limits_{n=-N}^{N-1}
\sqrt{A_nA_{n+1}}\cos(\alpha_n-\alpha_{n+1}),\\
H_{\text{kin}}^b &= -J\sum\limits_{n=-N}^{N-1}
\sqrt{B_nB_{n+1}}\cos(\beta_n-\beta_{n+1}),
\end{aligned}
\label{Ekin12}
\end{equation}
\begin{equation}
H_{\text{int}}^a=\frac{g}{2}\sum\limits_{n=-N}^NA_n^2,\quad
H_{\text{int}}^b=\frac{g}{2}\sum\limits_{n=-N}^NB_n^2,
\label{Eint12}
\end{equation}
\begin{equation}
 \rho_a\equiv \sum_n |a_n|^2,\quad
 \rho_b\equiv \sum_n|b_n|^2,
\end{equation}
and angular brackets denote ensemble averaging.
By definition, $h_{\text{kin}}^{a,b}$ and $h_{\text{int}}^{a,b}$ can be thought of as one-species densities
of the kinetic and interaction energies, respectively.
According to the data shown in Figure~\ref{fig-enr}, 
only in the case of the in-phase state $p=0$ the energy densities are equally distributed among the species.
In the cases of the checkboard states $p=0.5$ and $p=1$, the first species predominantly concentrates the interaction energy, while the second 
one has the larger kinetic energy. So, we can conditionally refer to them as ``heavy'' and
``light'' species, correspondingly.
The heavy species is more apt to soliton formation, while the light species is more responsible 
for wavepacket spreading. Rabi inter-species coupling tends to remove distinction
between the light and heavy species, 
at least, on statistical average, as in the case of $p=0$.

\subsection{Spatial separation of species}
\label{Separ}

\begin{figure}[h!tb]
\begin{center}
\includegraphics[width=0.6\textwidth]{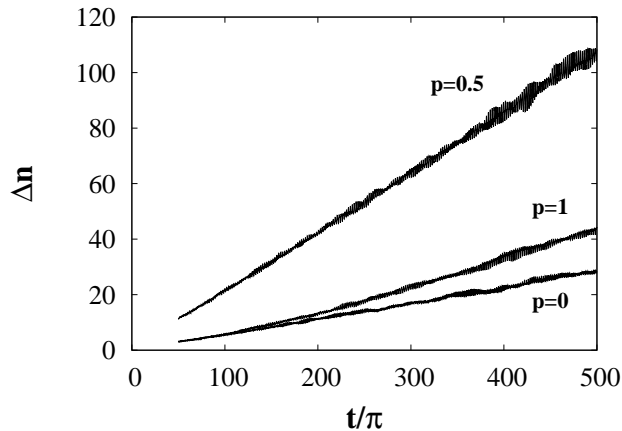}
\end{center}
\caption{Ensemble-averaged distance between species centers of mass vs time.
}
\label{fig-separ}
\end{figure}

It should be noted that the initial states (\ref{init}) for $p=0$, 0.5 and 1
are spatially symmetric. In the absence of random perturbations $\nu$ and $\xi$,
this symmetry is preserved in course of evolution, even in the regime of dynamical chaos.
It leads to the absence of center-of-mass motion.
Inclusion of random perturbations (\ref{nuxi}) violate the spatial symmetry and there appears
a possibility for mutual displacement of species, especially under
the miscible/immiscible crossover.
Let's denote distance between species centers of mass as
\begin{equation}
 \Delta n = \left<\left|n_{\text{a}} - n_{\text{b}}\right|\right>,
 \label{sep} 
\end{equation}
where 
\begin{equation}
 n_{\text{a}} = \frac{\sum_n n|a_n|^2}{\sum_n |a_n|^2},\quad
 n_{\text{b}} = \frac{\sum_n n|b_n|^2}{\sum_n |b_n|^2}.
\end{equation}
Figure \ref{fig-separ} shows that inter-species distance increases, on average, linearly with time,
implying motion with constant velocity. The velocity depends on the parameter $p$: 
it is the largest in the case of $p=0.5$, while $p=0$ corresponds to the smallest velocity.
Comparison of this dependence to the data presented in Fig.~\ref{fig-z2} 
indicates connection of the species separation to dynamics of Rabi inter-level oscillations.
Indeed, it is reasonable to suggest that inter-species transformation has to 
reduce mean inter-species distance.
So, the fastest separation is observed in the case of $p=0.5$, when Rabi oscillations are suppressed
by the internal self-trapping.

\section{Results and discussion} 
\label{Summ}

The present work is devoted to study of the binary discrete nonlinear Schr\"odinger equation (DNLSE)
describing dynamics of two-species Bose-Einstein condensate loaded into an optical lattice.
Different species are coupled to each other by external rf radiation causing
inter-species transitions.
Attention is concentrated on soliton formation and 
the miscible/immiscible crossover in the regime of strong nonlinearity,
corresponding to high condensate density and/or strong inter-atomic coupling.
We consider the case of comparable rates of tunneling and inter-species transitions, that is, 
these processes are competing with each other.
It is found that condensate dynamics qualitatively depends on configuration of the initial state.
The state being a sequence of occupied and unoccupied sites ($p=0.5$) exposes relatively stable dynamics
with strong impact of self-trapping in both spatial and internal degrees of freedom.
In contrast, initial states without unoccupied holes inside expose strong dynamical chaos revealing itself in
highly irregular behavior of spatial structure and internal state.
Onset of chaos is accompanied by drastic jump-like variations of the kinetic and interaction energies.
Such jumps suggest that the corresponding initial conditions are far from bound states.
So, one can conclude that stable solitonic configuration is possible only under
spatial separation of solitons, like in the case of $p=0.5$.

Strength of chaos can be quantified by analysing evolution of distance in the Hilbert space between two
initially close states. It allows one to define the corresponding Lyapunov exponent, as well as its finite-time
counterpart. Lyapunov analysis shows that, after some beginning time interval, chaos ceases and dynamics acquires stability.
Stabilization is accompanied by division of condensate onto localized and delocalized fractions.
Localized fraction represents some kind of a equillibrium state
being close to one of bound states. 
It consists of immiscible and spatially separated solitons.
However, neither the positions of solitons nor the species they are formed by are predictable.
Analysis by means of Monte-Carlo sampling shows that statistical properties 
of this equillibrium state also depend on a form of an initial state.
In particular, the initial state with the checkboard phase configuration ($p=1$) 
leads to equillibrium states with higher density and, therefore,
shows stronger tendency to spatial and internal self-trapping.
In contrast, the in-phase initial state ($p=0$) exposes stronger impact of Rabi inter-species
transitions that is reflected in zero ensemble-averaged population imbalance.

Phenomenon of the chaos-assisted soliton formation in Bose-Einstein condensates can be thought of as some
manifestation of self-organization. Here one should remind that self-organization basically occurs
in dissipative dynamical systems. 
Despite the discrete nonlinear Schr\"odinger equation has a Hamiltonian form,
this is a dynamical system with many degrees of freedom that can exhibit dissipative features.
In our particular case the important role is played by generation of the delocalized fraction.
Leaving the region of strong interaction, this fraction carries away some part of energy and thereby
facilitates the self-organization. In this way there arises a question: how long the self-organization can be 
affected by the presence of a confining potential that is always present in experiments?
Another issue deserving to be addressed is influence of effects which are not taken into account by
the mean-field aprroximation the DNLSE relies upon. In particular, chaos infers emergence of condensate 
excitations and generation of non-condensed fraction \cite{Burgdorfer-depletion}. 
These issues are worth from the viewpoint of experimental observation of the chaos-assisted soliton
formation. They will be studied in forthcoming works.

\subsection{Conclusions}

Main results of the work can be formulated as follows:
\begin{itemize}
 \item stable solitonic configuration consists of spatially separated immiscible solitons;
 \item onset of chaos can be accompanied by jumps of kinetic
 and interaction energies;
 \item solitonic pattern undergoes the spontaneous self-stabilization after emittance of ballistically propagating waves;
 \item the crossover to the self-stabilization and formation of stable immiscible solitons are reflected in remarkable lowering
 of the finite-time Lyapunov exponent, characterizing divergence of wave states in the Hilbert space.
\end{itemize}

\section*{Acknowledgements}

This work was partially supported by the Russian Foundation of Basic Research
within the projects 15-02-00463-a and 15-02-08774-a.
Authors are grateful to D.N.~Maksimov and A.R.~Kolovsky for
helpful comments concerning the subject of the research.












\end{document}